\documentclass[aos,MSNbibl,nameyear,seceqn,dvips]{arximspdf}
\usepackage{dcolumn}
\usepackage{graphicx,url}
\doi{10.1214/14-AOS1233} 
\volume{42}
\issue{5}
\pubyear{2014}
\firstpage{1751}
\lastpage{1786}

\makeatletter
\newcolumntype{d}[1]{D{.}{.}{#1}}

\newcommand{\rrvert}{\vert}
\newcommand{\llvert}{\vert}
\renewcommand{\citep}[1]{(\citeauthor{#1} \citeyear{#1})}
\newcommand{\eqref}[1]{(\ref{#1})}
\newproclaim{example}{Example}

\newproclaim{remark}{Remark}
\newtheorem{theorem}{Theorem}
\newtheorem{condition}{Condition}
\newtheorem{proposition}{Proposition}
\newtheorem{lemma}{Lemma}
\newproclaim{fact}{Fact}

\newproclaim{definition}{Definition}
\newproclaim{assumption}{Assumption}

\newcommand{\bx}{\mathbf{x}}

\newcommand{\cC}{\mathcal{C}}
\makeatother

\begin{document}
\begin{frontmatter}

\title{Variable selection for general index models via sliced inverse
regression}
\runtitle{SIR with variable selection}

\begin{aug}
\author{\fnms{Bo}~\snm{Jiang}\thanksref{t1}\ead[label=e1]{bojiang83@gmail.com}}
\and
\author{\fnms{Jun S.}~\snm{Liu}\corref{}\thanksref{t1,t2}\ead[label=e2]{jliu@stat.harvard.edu}}
\thankstext{t1}{Supported in part by NSF Grants DMS-10-07762 and DMS-11-20368.}

\thankstext{t2}{Supported in part by Shenzhen Special Fund for Strategic
Emerging Industry Grant ZD201111080127A while Jun S. Liu was a
Guest Professor at Tsinghua University in summers of 2012 and 2013.}
\runauthor{B.~Jiang and J.~S.~Liu}
\affiliation{Harvard University}
\address{Department of Statistics\\
Harvard University\\
1 Oxford Street\\
Cambridge, Massachusetts 02138\\
USA\\
\printead{e1}\\
\phantom{E-mail:\ }\printead*{e2}}\vspace*{-3pt} 
\end{aug}

\received{\smonth{4} \syear{2013}}
\revised{\smonth{4} \syear{2014}}

%
\begin{abstract}
Variable selection, also known as feature selection in machine
learning, plays an important role in modeling high
dimensional data and is key to data-driven scientific
discoveries. We consider here the problem of detecting influential
variables under the general index model, in which the
response is dependent of predictors through
an unknown function of one or more linear combinations of them.
Instead of building a predictive model of the
response given combinations of predictors, we model the
conditional distribution of predictors given the response.
This inverse modeling perspective motivates us to propose a stepwise procedure
based on likelihood-ratio tests, which is effective and computationally
efficient in identifying important variables without specifying a
parametric relationship between predictors and the
response. For example, the proposed procedure is able to detect
variables with pairwise, three-way or even higher-order interactions
among $p$ predictors with a computational time of $O(p)$ instead of
$O(p^k)$ (with $k$ being the highest order of interactions).
Its excellent empirical performance in
comparison with existing methods is demonstrated through
simulation studies as well as real data examples.
Consistency of the variable selection procedure when both the number of
predictors
and the sample size go to infinity is established.\vspace*{-3pt}
\end{abstract}

%
\end{frontmatter}

\section{Introduction}

Recently, there has been a significant surge of interest in
analytically accurate, numerically robust, and algorithmically
efficient variable selection methods, largely due to the tremendous
advance in data collection techniques such as those in biology, finance,
internet, etc. The importance of discovering truly
influential factors from a large pool of possibilities is now widely
recognized by both general scientists and quantitative modelers. Under
linear regression models, various regularization methods have been
proposed for simultaneously estimating regression coefficients and
selecting predictors. Many promising algorithms, such as Lasso
[\citet{Tibshirani1996,Zou2006,Friedman2007}], LARS [\citet{Efron2004}]
and smoothly clipped absolute deviation [SCAD; \citet{Fan2001}], have
been invented. When the number of the predictors is extremely large,
\citet{Fan2008}\vadjust{\goodbreak} have proposed a sure independence screening (SIS)
framework that first independently selects variables based on their
correlations with the response and then applies variable selection
methods.

\subsection{Sliced inverse regression with variable selection}\label{secsir}

When the relationship between the response $Y$ and predictors
$\mathbf{X} = (X_1,X_2,\ldots,X_p)^T$ is beyond linear, performances of
variable selection methods for linear models can be severely
compromised. In his seminal paper on dimension reduction,
\citet{Li1991} proposed a semiparametric index model of the form
%
\begin{equation}
\label{eqsir1} Y = f\bigl(\bolds\beta^T_1\mathbf{X},
\bolds\beta^T_2\mathbf{X},\ldots,\bolds
\beta^T_q\mathbf{X},\varepsilon\bigr),
\end{equation}
where $f$ is an unknown link function and $\varepsilon$ is a stochastic
error independent of $\mathbf{X}$, and the sliced inverse regression (SIR)
method to estimate the so-called
sufficient dimension reduction (SDR) directions
$\bolds\beta_1,\ldots,\bolds\beta_q$.

Given independent observations $\{(\bx_i,y_i)\}_{i=1}^n$, SIR
first divides the range of the $y_i$ into $H$ disjoint intervals,
denoted as $S_1,\ldots, S_H$, and computes for $h=1,\ldots, n$,
$\bx_h =n_h^{-1} \sum_{y_i \in S_h}\bx_i$, where $n_h$ is the number
of $y_i$'s in $S_h$. Then SIR estimates $\operatorname{Cov} (\mathbb
{E}(\mathbf{X}|Y) )$ by $\widehat{M}=n^{-1}\sum_{h=1}^H
n_h(\bx_h-\bar{\bx})(\bx_h-\bar{\bx})^T$ and $\operatorname{Cov} (\mathbf{X})$
by the
sample covariance matrix $\widehat{\Sigma}$. Finally, SIR uses the first
$K$ eigenvectors of $\widehat{\Sigma}^{-1} \widehat{M}$ to estimate
the SDR
directions, where $K$ is an estimate of $q$ based on the data.

For the ease of presentation, we assume that $\mathbf{X}$ has
been standardized such that $\mathbb{E} (\mathbf{X} ) = 0$ and
$\operatorname{Cov} (\mathbf{X} )=\mathbf{I}_p$. Eigenvalues of
$\operatorname{Cov} (\mathbb{E} (\mathbf{X}|Y ) )$ also
connects SIR with multiple linear regression (MLR). In MLR, the
correlation squared can be expressed as
\[
R^2 = \max_{\mathbf{b} \in\mathbb{R}^p} \bigl[\operatorname{Corr}
\bigl(Y,\mathbf {b}^T\mathbf{X} \bigr) \bigr]^2,
\]
while in SIR, the largest eigenvalue of $\operatorname{Cov} (\mathbb{E}
(\mathbf{X}|Y ) )$, called the first \emph{profile-$R^2$}, can
be defined as
\[
\lambda_1 \bigl(\operatorname{Cov} \bigl(\mathbb{E} (\mathbf{X}|Y )
\bigr) \bigr) = \max_{\mathbf{b} \in\mathbb{R}^p}\max_T \bigl[
\operatorname{Corr} \bigl(T(Y),\mathbf {b}^T\mathbf{X} \bigr)
\bigr]^2,
\]
where the maximization is taken over all bounded transformations
$T(\cdot)$ and vectors $\mathbf{b} \in\mathbb{R}^p$
[\citet{Chen1998}]. We can further define the $k$th profile-$R^2$,
$\lambda_k$ ($2 \leq k \leq q$), as the $k$th largest eigenvalue of
$\operatorname{Cov} (\mathbb{E} (\mathbf{X}|Y ) )$ by
restricting the vector $\mathbf{b}$ to be orthogonal to eigenvectors of
the first $(k-1)$ profile-$R^2$.

Since the estimation of SDR directions does not automatically lead to
variable selection, various methods have been developed to perform
dimension reduction and variable selection simultaneously for index
models. For example, \citet{Li2005} designed a backward subset selection
method based on $\chi^2$-tests derived in \citet{Cook2004}, and
\citet{Li2007} developed the sparse SIR (SSIR) algorithm to obtain
shrinkage estimates of the SDR directions under $L_1$ norm. Motived by
the F-test in stepwise regression and the connection between SIR and
MLR, \citet{Zhong2012} proposed a forward stepwise variable selection procedure
called correlation pursuit (COP) for index models.

By construction, however, the original SIR method only extracts
information from the first conditional moment, $\mathbb{E} (\mathbf
{X}|Y )$.\vadjust{\goodbreak} When the link function $f$ in (\ref{eqsir1}) is
symmetric along a direction, it will fail to recover this direction.
Similarly, aforementioned variable selection methods based on SIR will
miss important variables with interaction or other second-order
effects. For example, if $Y = X_1^2+X_2^2+\varepsilon$ or
$Y=X_1X_2+\varepsilon$, then the profile-$R^2$ between $Y$ and $X_1$ or
$X_2$ will always be $0$.

\subsection{Introducing SIRI for general index models}\label{secint}

\begin{figure}

\includegraphics{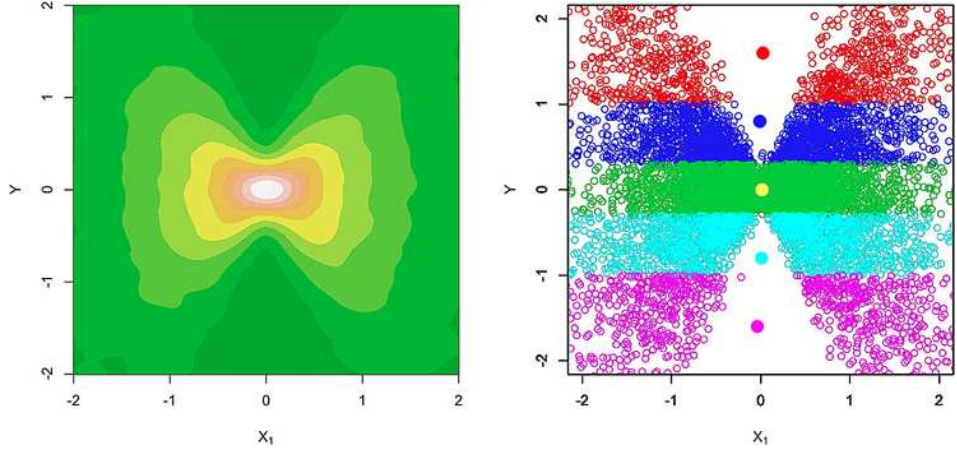}

\caption{Left panel: contour plot for the joint distribution of $Y$ and
$X_1$ in example (\protect\ref{example}). Right panel: conditional means (round
dots marks) of $X_1$ given slices of $Y$. Slices are indicated by
different colors. The corresponding conditional variances of $X_1$ are
(top to bottom): $2.29$, $0.92$, $0.41$, $0.98$ and $2.33$, respectively.}
\label{figsnote}
\end{figure}

Consider the following simple example with $p$ independent and normally
distributed predictor variables $\mathbf{ X} = (X_1,X_2,\ldots,X_p)^T$:
%
\begin{equation}
\label{example} Y=X_1X_2+\varepsilon,
\end{equation}
where $\mathbf{X} \sim\operatorname{MVN}_p(\mathbf{0},\mathbf{I}_p)$ and $\varepsilon
\sim N(0,0.1)$. Even if one knows that the true model is
a linear model with two-way interactions, one has to consider over
$p^2/2$ possible terms. Most existing variable selection methods
(including screening strategies)
can be too expensive to implement when one has a moderate number of
predictor variables, say $p=1000$. Moreover, without any knowledge of
the functional form, it is nearly impossible to do variable and
interaction detections in a forward regression setting. In this
article, we show that the inverse modeling perspective of SIR
complements well the forward regression approach and can be used to our
advantage in detecting complex relationships. As shown in Figure~\ref{figsnote}, however, the mean of $X_1$ (or $X_2$) conditional on
slicing is constant (i.e., 0). Thus, existing variable selection
methods based on classic SIR cannot detect $X_1$ or $X_2$ here, even
though conditional variances of $X_1$ (and $X_2$) are significantly
different across slices.
The following algorithm, SIR for variable selection via Inverse
modeling (henceforth, SIRI), which is the main focus of this article,
can find the true model with only $O(p)$ steps.

\textit{The SIRI algorithm}. Observations are $\{(\bx
_i,y_i)\}_{i=1}^n$, where $\bx_i $ is a $p$-dimen\-sional continuous
predictor vector and $y_i$ is a univariate response.
\begin{itemize}
\item We divide the range of $\{y_i\}_{i=1}^n$ into $H$ nonoverlapping
intervals (or ``slices'') $S_1,\ldots, S_H$, with $n_h$, the number of
observations in $S_h$, roughly the same for $h=1,\ldots, H$.
\item Let $\cC$ denote the set of predictors that have been selected
as relevant. Then, for a new candidate variable $X_j$ not in $\cC$, we compute
\[
\widehat{D}^*_{j|\mathcal{C}} = \log\widehat{\sigma}^2_{j|\mathcal{C}}-
\sum_{h=1}^H\frac{n_h}{n}\log \bigl[
\widehat{\sigma}^{(h)}_{j|\mathcal{C}} \bigr]^2,
\]
where $ [\widehat{\sigma}^{(h)}_{j|\mathcal{C}} ]^2$ is the
estimated error variance by regressing $X_j$ on $\mathbf{X}_{\mathcal
{C}}$ in
the $h$th slice, and $\widehat{\sigma}^2_{j|\mathcal{C}}$ is the
estimated error variance by regressing $X_j$ on $\mathbf{X}_{\mathcal{C}}$
using all the observations. Variable $X_j$ is added to $\cC$ if
$\widehat{D}^*_{j|\mathcal{C}}$ is sufficiently large, and ignored otherwise.
\item Each variable within $\cC$ is reexamined using the $\widehat
{D}^*$ statistic for possible removal.
\item The above two steps are repeated until no more variables can be
added to or removed from $\cC$.
\end{itemize}

Note that one always starts SIRI with $\cC=\varnothing$, in which case
$\widehat{D}^*_{j|\mathcal{C}}$ is reduced to a contrast of the
within-slice versus between-slice variances: $\widehat{D}^*_{j} = \log
\widehat{\sigma}^2_j-\sum_{h=1}^H  (n_h/n ) \log [\widehat
{\sigma}^{(h)}_j ]^2$.
This test statistic can be used as a sure independence screening
criterion when $p$ is extremely large to reduce the set of candidate
predictors. The full recursive SIRI procedure based $\widehat
{D}^*_{j|\mathcal{C}}$ can then be applied to the reduced set of variables.

To illustrate, we generated $200$ observations from example (\ref
{example}) and divided the range of $y$ into $5$ slices with $40$
observations in each slice, {that is}, $p=1000$, $H=5$, $n=200$
and $n_h=40$. We found that $ (n\widehat{D}^*_{1} ) = 62.48$
and $ (n\widehat{D}^*_{2} ) = 56.03$ are highly significant
compared with their null distributions, which will be shown to be
asymptotically $\chi^2(8)$ [empirically we observed that $\max_{j \in\{
3,4,\ldots,1000\}} (n\widehat{D}^*_{j} ) = 28.46$]. So both
$X_1$ and $X_2$ can be easily detected from the screening stage. We
also tested whether $X_2$ can be correctly selected conditioning on
$X_1$ by calculating $ (n\widehat{D}^*_{2|\{1\}} ) = 148.83$.
This is also highly significant compared to its null distribution,
which is asymptotically $\chi^2(12)$ [or to contrast with $\max_{j \in
\{3,4,\ldots,1000\}} (n\widehat{D}^*_{j|\{1\}} ) = 31.85$]. We
were thus able to detect both $X_1$ and $X_2$ with a computational
complexity of $O(p)$.

Note that our main goal here is to select relevant predictors without
explicitly stating analytic forms through which they influence $y$. We
leave the construction of a specific parametric form to downstream
analysis, which can be applied to a small number of selected
predictors. For example, to pinpoint the specific interaction term
$X_1X_2$ in example (\ref{example}), one can apply linear-model based
methods to an expanded set of predictors that includes multiplicative
interactions between selected variables $\{X_1,X_2\}$.

\subsection{Related work}

There has been considerable effort in fitting models with interactions
and other nonlinear effects in recent statistical literatures. For
example, \citet{Ravikumar2009} introduced SpAM (sparse additive
nonparametric regression model) that generalizes sparse linear models
to the additive, nonparametric setting. \citet{Bien2012} developed
hierNet, an extension of Lasso to consider interactions in a model if
one or both variables are marginally important (referred to as
hierarchical interactions by the authors). \citet{Li2012} proposed a
sure independence screening procedure based on distance correlation
(DC-SIS) that is shown to be capable of detecting important variables
when interactions are presented.

The inverse modeling perspective that motivates this paper has been
taken by several researchers and has led to new developments in
dimension reduction and variable selection methods. \citet{Cook2007}
proposed inverse regression models for dimension reduction, which have
deep connections with the SIR method. \citet{Simon2012} proposed a
permutation-based method for testing interactions by exploring the
connection between the forward logistic model and the inverse normal
mixture model when the response $Y$ is binary. Another classical method
derived from the inverse modeling perspective is the na{\" i}ve Bayes
classifier for classifications with high dimensional features. Although
Na{\"i}ve Bayes classifier is limited by its strong independence
assumption, it can be generalized by modeling the joint distribution of
features. \citet{Murphy2010} proposed a variable selection method using
Bayesian information criterion (BIC) for model-based discriminant
analysis. \citet{Zhang2007} proposed a Bayesian method called BEAM to
detect epistatic interactions in genome-wide case--control studies,
where $Y$ is binary and the $\mathbf{X}$ are discrete.

The rest of the article is organized as follows. At the beginning of
Section~\ref{secmodel}, we introduce an inverse model of predictors
given slices of response and explore its link with SIR. A
likelihood-ratio test statistic for selecting relevant predictors under
this model is derived in Section~\ref{seclrt}, which is shown to be
asymptotically equivalent to the COP statistic in \citet{Zhong2012}. We
augment the inverse model to detect predictors with second-order
effects in Section~\ref{secaug}. A sure independence screening
criterion based on the augmented model is proposed in Section~\ref{secsis}.
A few theoretical results regarding selection consistency of the
proposed methods are described in Section~\ref{sectheory}.
By cross-stitching independence screening and likelihood-ratio tests,
an iterative stepwise procedure that we referred to as SIRI is
developed in Section~\ref{secimp}. Various implementation issues
including the choices of slicing schemes and thresholds are also
discussed. Simulations and real data examples are reported in
Sections~\ref{secsimulation} and~\ref{secreal}. Additional remarks in
Section~\ref{secconclusion} conclude the paper. Proofs of the theorems
are provided in the \hyperref[app]{Appendix}.

\section{Variable selection via a sliced inverse model}\label{secmodel}

Let $Y \in\mathbb{R}$ be a univariate response variable and $\mathbf
{X}=(X_1,X_2,\ldots,X_p)^T \in\mathbb{R}^p$ be a vector of $p$
continuous predictor variables. Let $\{(\bx_i, y_i)\}_{i=1}^{n}$ denote
$n$ independent observations on $(\mathbf{X}, Y)$. For discrete
responses, the $y_i$'s can be naturally grouped into a finite number of
classes. For continuous responses, we divide the range of $\{y_i\}
_{i=1}^n$ into $H$ disjoint intervals $S_1,\ldots,S_H$, also known as
``slices.'' Let $S(Y)$ indicate the slice membership of response $Y$,
that is, $S(Y) = h$ if $Y \in S_h$. For a fixed slicing scheme, we
denote $n_h=|S_h| \equiv n s_h$ where $\sum_{h=1}^Hs_h = 1$.

To view SIR from a likelihood perspective, we start with a seemingly
different model. We assume that the distribution of predictors given
the sliced response is multivariate normal:
%
\begin{equation}
\label{eqinv} \mathbf{X} | Y \in S_h \sim\operatorname{MVN} (
\mu_h, \Sigma ),\qquad 1 \leq h \leq H,
\end{equation}
where $\mu_h \in\mu+ \mathbb{V}^q$ belongs to a $q$-dimensional
affine space, $\mathbb{V}^q$ is a $q$-dimen\-sional subspace ($q<p$) and
$\mu\in\mathbb{R}^p$. Alternatively, we can write $\mu_h = \mu+
\Gamma\gamma_h$, where $\gamma_h \in\mathbb{R}^q$ and $\Gamma$ is a
$p$ by $q$ matrix whose columns form a basis of the subspace $\mathbb
{V}^q$. Although this representation is only unique up to an orthogonal
transformation on the bases $\Gamma$, the subspace $\mathbb{V}^q$ is
unique and identifiable. The following proposition proved by \citet
{Szretter2009} links the inverse model (\ref{eqinv}) with SIR.
%
\begin{proposition}
\label{propsir}
The maximum likelihood estimate (MLE) of the subspace $\mathbb{V}^q$ in
model (\ref{eqinv}) coincides with the subspace spanned by SDR
directions estimated from the SIR algorithm.
\end{proposition}


\subsection{Likelihood-ratio tests for detecting variables with mean
effects}\label{seclrt}

For the purpose of variable selection, we partition predictors into two
subsets: a set of relevant predictors indexed by $\mathcal{A}$ and a
set of redundant predictors indexed by $\mathcal{A}^c$, and assume the
following model:
%
\begin{eqnarray}
\label{eqsim} \mathbf{X}_{\mathcal{A}} | Y \in S_h &\sim&
\operatorname{MVN} \bigl(\mu_h \in \mu+ \mathbb{V}^q,
\Sigma \bigr),
\nonumber
\\[-8pt]
\\[-8pt]
\nonumber
\mathbf{X}_{\mathcal{A}^c} | \mathbf{X}_{\mathcal{A}}, Y \in
S_h &\sim& \operatorname{MVN} \bigl(\alpha+\bolds\beta^T
\mathbf{X}_{\mathcal{A}}, \Sigma _0 \bigr).
\end{eqnarray}
That is, we assume that the conditional distribution of relevant
predictors follows the inverse model (\ref{eqinv}) of SIR and has a
common covariance matrix in different slices. Given the current set of
selected predictors indexed by $\mathcal{C}$ with dimension $d$ and
another predictor indexed by $j \notin\mathcal{C}$, we propose the
following hypotheses:
\[
H_0\dvtx \mathcal{A} = \mathcal{C} \quad\mbox{v.s.}\quad H_1
\dvtx \mathcal{A} = \mathcal {C}\cup\{j\}.
\]
Let $L_{j|\mathcal{C}}$ denote the likelihood-ratio test statistic for
testing $H_1$ against $H_0$. In \citet{Jiang2014}, we showed that the
scaled log-likelihood-ratio test statistic is given by
%
\begin{equation}
\label{eqhomo} \widehat{D}_{j|\mathcal{C}} = \frac{2}{n}\log
(L_{j|\mathcal
{C}} ) = \sum_{k=1}^q\log
\biggl(1+{\frac{\widehat{\lambda}^{d+1}_k-\widehat
{\lambda}^d_k}{1-\widehat{\lambda}^{d+1}_k}} \biggr),
\end{equation}
where $\widehat{\lambda}^d_k$ and $\widehat{\lambda}^{d+1}_k$ are
estimates of the $k$th profile-$R^2$ based on $\bx_{\mathcal{C}}$ and
$\bx_{\mathcal{C}\cup\{j\}}$, respectively.
Since
${\frac{\widehat{\lambda}^{d+1}_k-\widehat{\lambda}^d_k}{1-\widehat
{\lambda}^{d+1}_k}}\mathop{\rightarrow}\limits^{P} 0$ as
$n \rightarrow\infty$ under
$H_0$ and that $\log(1+t) = t + O(t^2)$, we have
\[
2\log (L_{j|\mathcal{C}} ) = (n\widehat{D}_{j|\mathcal
{C}} ) = n\sum
_{k=1}^q \frac{\widehat{\lambda}^{d+1}_k-\widehat
{\lambda}^{d}_k}{1-\widehat{\lambda}^{d+1}_k}+o_p(1)
\mathop{\rightarrow}\limits^{d} \chi^2(q).
\]
This expression coincides with the COP statistics proposed by \citet
{Zhong2012}, which are defined as
\[
\operatorname{COP}^{d+1}_k = n \frac{\widehat{\lambda}^{d+1}_k-\widehat{\lambda
}^{d}_k}{1-\widehat{\lambda}^{d+1}_k},\qquad k=1,2,
\ldots,q \quad\mbox{and}\quad \operatorname{COP}^{d+1}_{1\dvtx q} = \sum
_{k=1}^q \operatorname{COP}^{d+1}_k.
\]
For all the predictors\vspace*{1pt} indexed by $j \in\mathcal{C}^c$, we can also
obtain the asymptotic joint distribution of $2\log (L_{j|\mathcal
{C}} ) =  (n\widehat{D}_{j|\mathcal{C}} )$ under the null
hypothesis with fixed number of predictors $p$ and as $n \rightarrow
\infty$,
%
\begin{equation}
\label{eqdist1} 2\log (L_{j|\mathcal{C}} ) = (n\widehat{D}_{j|\mathcal
{C}}
)_{j \in\mathcal{C}^c} \mathop{\rightarrow}\limits^{d} \Biggl(\sum_{k=1}^q
z^2_{kj} \Biggr)_{j \in\mathcal{C}^c},
\end{equation}
where $\mathbf{z}_k =  (z_{kj} )_{j \in\mathcal{C}^c} \sim\operatorname
{MVN} (\mathbf{0}, [\operatorname{Corr} (X_i,X_j|\mathbf{X}_{\mathcal
{C}} ) ]_{i, j \in\mathcal{C}^c} )$ and $\mathbf{z}_k$'s
are independent. Furthermore, we can show that, as $n \rightarrow\infty$,
\begin{eqnarray*}
\widehat{D}_{j|\mathcal{C}} &\mathop{\rightarrow}\limits^{\mathrm{a.s.}}& D_{j|\mathcal{C}}
\\
&=& \log \biggl(1+ \frac{{\operatorname{Var}(M_j)-\operatorname{Cov} (M_j,\mathbf{X}_{\mathcal{C}}
) [\operatorname{Cov} (\mathbf{X}_{\mathcal{C}} )
 ]^{-1}\operatorname{Cov} (M_j,\mathbf{X}_{\mathcal{C}}
)^T}}{\mathbb{E} (V_j )} \biggr),
\end{eqnarray*}
where $M_j = \mathbb{E} (X_j|\mathbf{X}_{\mathcal{C}},S(Y) )$,
$V_j = \operatorname{Var} (X_j|\mathbf{X}_{\mathcal{C}},S(Y) )$ and
$S(Y) = h$ when $Y \in S_h$ ($1 \leq h \leq H)$. 
By the Cauchy--Schwarz inequality and the normality assumption,
\[
D_{j|\mathcal{C}} = 0 \quad\mathrm{iff}\quad \mathbb{E} (X_j|\mathbf
{X}_{\mathcal{C}},Y \in S_h ) = \mathbb{E} (X_j|
\mathbf {X}_{\mathcal{C}} ),\qquad 1 \leq h \leq H.
\]
That is, the test statistic $\widehat{D}_{j|\mathcal{C}}$ almost surely
converges to zero if the conditional mean of $X_j$ is independent of
slice membership $S(Y)$. See \citet{Jiang2014} for detailed proofs about
properties of $\widehat{D}_{j|\mathcal{C}}$.

Given thresholds $\nu_a>\nu_d$ and the current set of selected
predictors indexed by~$\mathcal{C}$, we can select relevant variables
by iterating the following steps until no new addition or deletion occurs:
\begin{itemize}
\item Addition step: find $j_a$ such that $\widehat{D}_{j_a|\mathcal
{C}} = \max_{j \in\mathcal{C}^c} \widehat{D}_{j|\mathcal{C}}$; let
$\mathcal{C} = \mathcal{C}+\{j_a\}$ if $\widehat{D}_{j_a|\mathcal{C}} >
\nu_a$.
\item Deletion step: find $j_d$ such that $\widehat{D}_{j_d|\mathcal
{C}-\{j_d\}} = \min_{j \in\mathcal{C}} \widehat{D}_{j|\mathcal{C}-\{j\}
}$; let $\mathcal{C} = \mathcal{C}-\{j_d\}$ if $\widehat
{D}_{j_d|\mathcal{C}-\{j_d\}} < \nu_d$.
\end{itemize}

In Section~\ref{secthr1}, we will study the selection consistency of
the foregoing procedure under model (\ref{eqsim}), allowing for the
number of predictors $p$ to grow with the sample size $n$. 

\subsection{Detecting variables with second-order effects}
\label{secaug}

Let us revisit example~(\ref{example}).
As illustrated in Figure~\ref{figsnote}, we have $\mathbb{E}
(X_j|Y \in S_h ) = 0$ for $j=1,2$ and $1\leq h \leq H$. Starting
with $\mathcal{C}=\varnothing$, the stepwise procedure in Section~\ref{seclrt}
fails to capture either $X_1$ or $X_2$ since $D_{1|\mathcal{C}=\varnothing
} = D_{2|\mathcal{C}=\varnothing} = 0$. In order to detect predictors
with different (conditional) variances across slices, such as $X_1$ and
$X_2$ in this example, we augment model (\ref{eqsim}) to a more
general form,
%
\begin{eqnarray}
\label{eqaug} \mathbf{X}_{\mathcal{A}} | Y \in S_h &\sim&
\operatorname{MVN} (\mu_h, \Sigma _h ),
\nonumber
\\[-8pt]
\\[-8pt]
\nonumber
\mathbf{X}_{\mathcal{A}^c} | \mathbf{X}_{\mathcal{A}}, Y \in
S_h &\sim& \operatorname{MVN} \bigl(\alpha+\bolds\beta^T
\mathbf{X}_{\mathcal{A}}, \Sigma _0 \bigr),
\end{eqnarray}
which differs from model (\ref{eqsim}) in its allowing for
slice-dependent means and covariance matrices for relevant predictors.
To guarantee identifiability, variables indexed by $\mathcal{A}$ in
model (\ref{eqaug}) have to be minimally relevant, that is, $\mathcal
{A}$ does not contain any predictor that is conditionally independent
of $Y$ given the remaining predictors in $\mathcal{A}$. \citet
{Jiang2014} gave a rigorous proof of the uniqueness of minimally
relevant predictor set $\mathcal{A}$.\vadjust{\goodbreak}

By following the same hypothesis testing framework as in Section~\ref{seclrt}, we can derive the scaled log-likelihood-ratio test statistic
under the augmented model~(\ref{eqaug}):
%
\begin{equation}
\label{eqheter} \widehat{D}^*_{j|\mathcal{C}} = \log\widehat{\sigma}^2_{j|\mathcal{C}}-
\sum_{h=1}^H\frac{n_h}{n}\log \bigl[
\widehat{\sigma}^{(h)}_{j|\mathcal{C}} \bigr]^2,
\end{equation}
where $\mathcal{C}$ is the set of currently selected predictors and $j
\in
\mathcal{C}^c$,
$ [\widehat{\sigma}^{(h)}_{j|\mathcal{C}} ]^2$ is the
estimated variance by regressing $X_j$ on $\mathbf{X}_{\mathcal{C}}$ in
slice $S_h$, and $\widehat{\sigma}^2_{j|\mathcal{C}}$ is the
estimated variance by regressing $X_j$ on $\mathbf{X}_{\mathcal{C}}$
using all the observations. Although model (\ref{eqaug})
involves more parameters than model (\ref{eqsim}), by relaxing the
homoscedastic constraint on the distribution of relevant predictors
across slices, the form of the likelihood-ratio test statistic in
(\ref{eqheter}) appears simpler than that in (\ref{eqhomo}). The
augmented test statistic $ (n\widehat{D}^*_{j|\mathcal{C}} )$
was used to select relevant predictors in the illustrative example of
Section~\ref{secint}.

Under the assumption that $\mathcal{A} \subset\mathcal{C}$ with
$|\mathcal{C}| = d$, we can derive the exact and asymptotic
distribution of $ (n\widehat{D}^*_{j|\mathcal{C}} )$:
\begin{eqnarray*}
n\widehat{D}^*_{j|\mathcal{C}} &\sim& n\log \biggl(1+\frac{Q_0}{\sum_{h=1}^H Q_h}
\biggr) - \sum_{h=1}^H \frac{n_h}{n}
\log \biggl( \frac
{Q_h/n_h}{\sum_{h=1}^HQ_h / n} \biggr)
\\
&\mathop{\rightarrow}\limits^{d}& \chi^2\bigl((H-1) (d+2)\bigr),
\end{eqnarray*}
where $Q_0 \sim\chi^2((H-1)(d+1))$ and $Q_h \sim\chi^2(n_h-(d+1))$
($1 \leq h \leq H$) are mutually independent according to \emph
{Cochran's theorem}.
For all the predictors indexed by $j \in\mathcal{C}^c$, we can also
obtain the asymptotic joint distribution of $ (n\widehat
{D}^*_{j|\mathcal{C}} )$ under the assumption that $\mathcal{A}
\subset\mathcal{C}$ (with $p$ fixed and $n\rightarrow\infty$):
%
\begin{equation}
\label{eqdist2} \bigl(n\widehat{D}^*_{j|\mathcal{C}} \bigr)_{j \in\mathcal{C}^c}
\mathop{\rightarrow}\limits^{d} \Biggl(\sum_{i=1}^{(H-1)(d+1)}z^2_{ij}+
\sum_{i=1}^{H-1}\widetilde{z}^2_{ij}
\Biggr)_{j \in\mathcal{C}^c},
\end{equation}
where $\mathbf{z}_i$'s and $\widetilde{\mathbf{z}}_i$'s are mutually
independent with
\[
\mathbf{z}_i = (z_{ij} )_{j \in\mathcal{C}^c} \sim\operatorname
{MVN} \bigl(\mathbf{0}, \bigl[\operatorname{Corr} (X_j,X_k|
\mathbf{X}_{\mathcal
{C}} ) \bigr]_{j, k \in\mathcal{C}^c} \bigr)
\]
and
\[
\widetilde{\mathbf{z}}_i = (\widetilde{z}_{ij}
)_{j \in\mathcal
{C}^c} \sim\operatorname{MVN} \bigl(\mathbf{0}, \bigl[
\operatorname{Corr}^2 (X_j,X_k|
\mathbf{X}_{\mathcal{C}} ) \bigr]_{j, k \in\mathcal
{C}^c} \bigr).
\]
When the number of predictors $p$ is fixed and the sample size $n
\rightarrow\infty$,
\begin{eqnarray*}
\widehat{D}^*_{j|\mathcal{C}} &\mathop{\rightarrow}\limits^{\mathrm{a.s.}}& D^*_{j|\mathcal
{C}}
\\
&=& \log \biggl(1+ \frac{{\operatorname{Var}(M_j)-\operatorname{Cov} (M_j,\mathbf{X}_{\mathcal{C}}
) [\operatorname{Cov} (\mathbf{X}_{\mathcal{C}} )
 ]^{-1}\operatorname{Cov} (M_j,\mathbf{X}_{\mathcal{C}}
)^T}}{\mathbb{E} (V_j )} \biggr)
\\
\nonumber
&&{}+\log\mathbb{E} (V_j ) - \mathbb{E}\log
(V_j ),
\end{eqnarray*}
where $M_j = \mathbb{E} (X_j|\mathbf{X}_{\mathcal{C}},S(Y) )$,
$V_j = \operatorname{Var} (X_j|\mathbf{X}_{\mathcal{C}},S(Y) )$ and
$S(Y) = h$ when $Y \in S_h$ ($1 \leq h \leq H$). According to the
Cauchy--Schwarz inequality and Jensen's inequality,
\begin{eqnarray*}
D^*_{j|\mathcal{C}} = 0 &&\quad\mbox{iff}\quad \mathbb{E} (X_j|
\mathbf {X}_{\mathcal{C}},Y \in S_h ) = \mathbb{E}
(X_j|\mathbf {X}_{\mathcal{C}} )\quad \mbox{and}\\
& &\qquad \operatorname{Var} (X_j|\mathbf{X}_{\mathcal{C}},Y
\in S_h ) = \operatorname {Var} (X_j|
\mathbf{X}_{\mathcal{C}} ),
\end{eqnarray*}
for $1 \leq h \leq H$. That is, the augmented test statistic $\widehat
{D}^*_{j|\mathcal{C}}$ almost surely converges to zero if both the
conditional mean and the conditional variance of $X_j$ is independent
of slice membership $S(Y)$. Detailed proofs of these properties are
collected in \citet{Jiang2014}.

A forward-addition backward-deletion algorithm similar to the stepwise
procedure proposed in Section~\ref{seclrt} can be used with the
augmented likelihood-ratio test statistic $\widehat{D}^*_{j|\mathcal
{C}}$. In Section~\ref{secthr2}, we will provide theoretical results
on the selection consistency of stepwise procedure based on $\widehat
{D}^*_{j|\mathcal{C}}$.

\subsection{Sure independence screening strategy: \texorpdfstring{SIS$^{\ast}$}{SIS*}}\label{secsis}
When dimensionality $p$ is very large, the performance of the stepwise procedure
can be compromised. We
recommend adding an independence screening step to first reduce the
dimensionality from ultra-high to moderately high. A natural choice of
the test statistic for the independence screening procedure is
$\widehat{D}^*_{j|\mathcal{C}}$ with $\mathcal{C}=\varnothing$, that is,
\[
\widehat{D}^*_{j} = \log\widehat{\sigma}^2_{j}-
\sum_{h=1}^H\frac{n_h}{n}\log \bigl[
\widehat{\sigma}^{(h)}_{j} \bigr]^2,
\]
where $ [\widehat{\sigma}^{(h)}_{j} ]^2$ is the estimated
variance of $X_j$ in slice $S_h$, and $\widehat{\sigma}^2_{j}$ is the
estimated variance of $X_j$ using all the observations. In Section~\ref{secthr3}, we will show that if we rank predictors according to $\{
\widehat{D}^*_j, 1\leq j \leq p\}$, then the sure independence
screening procedure, which we call SIS$^*$, that takes the first $o(n)$
predictors has a high probability (almost surely) of including relevant
predictors that have either different means or different variances
across slices.

\section{Theoretical results}\label{sectheory}
We here establish the selection consistency for procedures introduced
in Sections~\ref{seclrt} and~\ref{secaug}, as well as the
SIS$^*$ screening strategy in Section~\ref{secsis}.

\subsection{Selection consistency under homoscedastic model}\label{secthr1}
To proceed, we need the following concept to study the detectability of
relevant predictors under model~(\ref{eqsim}).
%
\begin{definition}[(First-order detectable)]\label{defmar} We say a
collection of predictors indexed by $\mathcal{C}_0$ is first-order
detectable if there exist $\kappa\geq0$ and\vadjust{\goodbreak} $\xi_0>0$ such that for
any set of predictors indexed by $\mathcal{C}$ and $\mathcal{C}^c \cap
\mathcal{C}_0 \neq\varnothing$,
\[
\max_{j \in\mathcal{C}^c \cap\mathcal{C}_0} \biggl[ \frac{\operatorname
{Var} (M_j )-\operatorname{Cov} (M_j,\mathbf{X}_{\mathcal{C}}
) [\operatorname{Cov} (\mathbf{X}_{\mathcal{C}} ) ]^{-1}\operatorname
{Cov} (M_j,\mathbf{X}_{\mathcal{C}} )^T}{\mathbb{E}
(V_j )} \biggr] \geq
\xi_0 n^{-\kappa},
\]
where
$M_j=\mathbb{E} (X_j|\mathbf{X}_{\mathcal{C}},S(Y) )$ and
$V_j=\operatorname{Var} (X_j|\mathbf{X}_{\mathcal{C}},S(Y) )$.
\end{definition}
 In the above definition, we allow the distribution of the
random samples $ (\mathbf{X},Y )$
to be dependent on the sample size $n$. For any first-order detectable
predictor, its conditional means given other predictors and different
slices are not all identical and differences among these conditional
means are not too small relative to the sample size. The
following example illustrates the implication of Definition~\ref{defmar}.
%
\begin{example}\label{exex1}
Suppose $Y$ is divided into two slices and there are two
predictors $ (X_1,X_2 )$. Conditional distributions of the $X$
given the slices are
\begin{eqnarray*}
\pmatrix{ X_1
\vspace*{2pt}\cr
X_2 }\bigg \vert Y \in S_1 &\sim&
\operatorname{MVN} \biggl( \pmatrix{ 1
\vspace*{2pt}\cr
1}, \pmatrix{ 1 & 1
\vspace*{2pt}\cr
1 & 2} \biggr)\quad\mbox{and}
\\
\pmatrix{ X_1
\vspace*{2pt}\cr
X_2 }\bigg\vert Y \in S_2 &\sim&
\operatorname{MVN} \biggl( \pmatrix{
 -1
\vspace*{2pt}\cr
-1 },\pmatrix{
\sigma^2 & \sigma^2
\vspace*{2pt}\cr
\sigma^2 & 2\sigma^2 }
\biggr).
\end{eqnarray*}
It is easy to show that $X_1$ is first-order detectable but
$X_2$ is
not because $\mathbb{E} (X_2 | X_1,Y\in S_h ) = X_1$, which
is identical for $h=1,2$. If $\sigma^2=1$, $X_2$ and $Y$ are
conditionally independent given $X_1$, and $X_2$ is indeed redundant
for predicting $Y$ if we
have already included $X_1$. If $\sigma^2 \neq1$, however,
$\operatorname{Var} (X_2|X_1, Y \in S_h )$ depends on $h$, and thus, $X_2$
is relevant for predicting $Y$ even if we have included~$X_1$. However, procedures that can only detect first-order
detectable predictors will miss $X_2$ in this case. 
\end{example}

Suppose the following conditions hold for predictors $\mathbf{X}$ with
dimension $p$.

\begin{condition}\label{cond1}
There exist $0 < \tau_{\mathrm{min}} < \tau_{\mathrm{max}} < \infty$ such that
\[
\tau_{\mathrm{min}}\leq\lambda_{\mathrm{min}} \bigl(\operatorname{Cov} (\mathbf {X}|Y
\in S_h ) \bigr)<\lambda_{\mathrm{max}} \bigl(\operatorname{Cov} (
\mathbf{X}|Y \in S_h ) \bigr)\leq\tau_{\mathrm{max}},
\]
and that
\[
\lambda_{\mathrm{max}} \bigl(\operatorname{Cov} ( \mathbf{X} ) \bigr)\leq
\tau_{\mathrm{max}},
\]
where $\lambda_{\mathrm{min}} ( \cdot )$ and
$\lambda_{\mathrm{max}} ( \cdot )$ denote the smallest and
largest eigenvalues, respectively, of a positive definite matrix.
\end{condition}
%
\begin{condition}\label{cond2}
$p = O(n^{\rho})$ as $n \rightarrow\infty$ with $\rho> 0$ and $2\rho
+ 2\kappa< 1$, where $\kappa$ is the same constant as in
Definition~\ref{defmar}.
\end{condition}
 Condition~\ref{cond1} excludes singular cases when some
predictors are constants or highly correlated. Assuming that\vadjust{\goodbreak}
Condition~\ref{cond1} holds, \citet{Jiang2014} gave an equivalent
characterization of first-order detectable predictors under model (\ref
{eqsim}). Condition~\ref{cond2}
allows the number of predictors $p$ to grow with the sample size $n$
but the growth rate cannot exceed $n^{{1}/{2}-\kappa}$. In
situations when $p$ is larger than $n^{{1}/{2}-\kappa}$, we can
first use the screening strategy SIS$^*$ introduced in Section~\ref{secsis} to reduce the dimensionality. In Section~\ref{secthr3}, we
will show theoretically that SIS$^*$ can be used to deal with scenarios
when $p$ is much larger than $n$. The following theorem, which is
proved in Appendix~\ref{app2}, guarantees that the
stepwise procedure described in Section~\ref{seclrt} is selection
consistent for
first-order detectable predictors if two thresholds $\nu_a$
and $\nu_d$ are chosen appropriately.

\begin{theorem}\label{thmsim}
Under model (\ref{eqsim}), Conditions~\ref{cond1} and~\ref{cond2}, if
the set of relevant predictors indexed by $\mathcal{A}$ is first-order
detectable with constant $\kappa$, then there exists constant $c>0$
such that
\begin{eqnarray*}\label{eqforward}
 & &\operatorname{Pr} \Bigl(\min_{\mathcal{C}\dvtx \mathcal{C}^c\cap\mathcal{A}
\neq\varnothing} \max
_{j \in\mathcal{C}^c} \widehat{D}_{j|\mathcal{C}} \geq cn^{-\kappa} \Bigr)
\\
&&\qquad\geq 1-O \biggl(p(p+1)q\exp{ \biggl(-N_1
\frac{n^{1-2\kappa}}{p^2q^2} \biggr)} \biggr) \rightarrow1
\end{eqnarray*}
and
\begin{eqnarray*}
\label{eqbackward} & &\operatorname{Pr} \biggl(\max_{\mathcal{C}\dvtx \mathcal{C}^c\cap\mathcal{A} =
\varnothing} \max
_{j \in\mathcal{C}^c} \widehat{D}_{j|\mathcal{C}} < \frac{c}{2}n^{-\kappa}
\biggr)
\\
&&\qquad\geq 1-O \biggl(p(p+1)q\exp{ \biggl(-N_2
\frac{n^{1-2\kappa
}}{p^2q^2} \biggr)} \biggr) \rightarrow1,
\end{eqnarray*}
as $n \rightarrow\infty$, where $N_1$ and $N_2$ are positive constants.
\end{theorem}
 The first convergence result implies that as long as the set
of currently selected predictors $\mathcal{C}$ does not contain all
relevant predictors in $\mathcal{A}$, {that is},
$\mathcal{C}^c\cap\mathcal{A} \neq\varnothing$, with probability going
to $1$ ($n \rightarrow\infty$) we can find a predictor $j \in
\mathcal{C}^c$ such that the test statistic
$\widehat{D}_{j|\mathcal{C}} \geq cn^{-\kappa}$. Thus, if we choose
the threshold $\nu_a = c n^{-\kappa}$ in the stepwise procedure, the
addition step will not stop selecting variables
until all relevant predictors have been included. On the other hand,
once all relevant predictors have been included in $\mathcal{C}$,
that is, $\mathcal{C}^c\cap\mathcal{A} = \varnothing$, the second
result guarantees that, with probability going to $1$,
$\widehat{D}_{j|\mathcal{C}} < (c/2)n^{-\kappa}<\nu_a$ for any
predictor $j \in\mathcal{C}^c$. Thus, the addition step will stop selecting
other predictors into $\mathcal{C}$. Consequently, if we choose
$\nu_d = (c/2)n^{-\kappa}$ in the deletion step, then all
redundant variables will be removed from the set of selected
variables until $\mathcal{C} = \mathcal{A}$ as $n\rightarrow\infty$.

\subsection{Selection consistency under augmented model}\label{secthr2}

Under model (\ref{eqaug}), we can further extend the definition of
detectability to include predictors with interactions and other
second-order effects.\vadjust{\goodbreak}
%
\begin{definition}[(Second-order detectable)]\label{defcond}
We call a collection of predictors indexed by $\mathcal{C}_2$
second-order detectable given predictors indexed by $\mathcal{C}_1$ if
$\mathcal{C}_2 \cap\mathcal{C}_1 = \varnothing$, and for any set
$\mathcal{C}$ satisfying $\mathcal{C}_1 \subset\mathcal{C}$ and
$\mathcal{C}_2 \not\subset\mathcal{C}$, there exist constants $\xi_1,
\xi_2>0$ and $\kappa\geq0$ such that either
%
\begin{equation}
\label{dt1} \max_{j \in\mathcal{C}^c \cap\mathcal{C}_2} \biggl[\frac{\operatorname
{Var} (M_j )-\operatorname{Cov} (M_j,\mathbf{X}_{\mathcal{C}}
) [\operatorname{Cov} (\mathbf{X}_{\mathcal{C}} ) ]^{-1}\operatorname
{Cov} (M_j,\mathbf{X}_{\mathcal{C}} )^T}{\mathbb{E}
(V_j )} \biggr]
\geq\xi_1 n^{-\kappa},\hspace*{-35pt}
\end{equation}
or
\[
\max_{j \in\mathcal{C}^c \cap\mathcal{C}_2} \bigl[\log (\mathbb {E} V_j ) -
\mathbb{E}\log (V_j ) \bigr] \geq\xi_2 n^{-\kappa},
\]
where $M_j = \mathbb{E} (X_j|\mathbf{X}_{\mathcal{C}},S(Y) )$,
$V_j = \operatorname{Var} (X_j|\mathbf{X}_{\mathcal{C}},S(Y) )$.
\end{definition}
 In other words, if the current selection $\mathcal{C}$
contains $\mathcal{C}_1$, then there always exist detectable predictors
conditioning on currently selected variables until we include all the
predictors indexed by $\mathcal{C}_2$. A relevant predictor $X_j$
indexed by $j \notin\mathcal{C}_2$ is \emph{not} second-order
detectable given $\mathcal{C}_1$ either because it is highly correlated
with some other predictors, or its effect can only be detected when
conditioning on predictors that have not been included in $\mathcal
{C}_1$. Based on Definition~\ref{defcond}, we define stepwise
detectable as follows.
%
\begin{definition}[(Stepwise detectable)]\label{defstep}
A collection of predictors indexed by $\mathcal{T}_0$ is said to be
$0$-stage detectable if $\mathbf{X}_{\mathcal{T}_0}$ is second-order
detectable conditioning on an empty set, and a collection of predictors
indexed by $\mathcal{T}_m$ is said to be $m$-stage detectable ($m \geq
1$) if $\mathbf{X}_{\mathcal{T}_m}$ is second-order detectable given
predictors indexed by $\bigcup_{i=1}^{m-1}\mathcal{T}_i$. Finally, a
predictor indexed by $j$ is said to be stepwise detectable if $j \in
\bigcup_{i=1}^{\infty}\mathcal{T}_i$.
\end{definition}
 According to Definition~\ref{defmar}, given the same
constant $\kappa$, there exists $\xi_1$ such that the set of
first-order detectable predictors defined in Definition~\ref{defmar}
is contained in the set of stepwise detectable predictors. The
following simple example illustrates the usefulness of foregoing definitions.
%
\begin{example}\label{exex2}
Suppose $Y$ is divided into two slices and there are only two
predictors $ (X_1,X_2 )$. Conditional distributions given the
slices are
\begin{eqnarray*}
 \pmatrix{ X_1
\vspace*{2pt}\cr
X_2 }
\bigg\vert Y \in S_1 &\sim&
\operatorname{MVN} \biggl( \pmatrix{ 0
\vspace*{2pt}\cr
0}, \pmatrix{
\sigma_1^2 & 1
\vspace*{2pt}\cr
1 & 1 }
\biggr)\quad\mbox{and}
\\
\pmatrix{ X_1
\vspace*{2pt}\cr
X_2 }
\bigg\vert Y \in S_2 &\sim&
\operatorname{MVN} \biggl(\pmatrix{ 0
\vspace*{2pt}\cr
0 }, \pmatrix{
\sigma_2^2 & -1
\vspace*{2pt}\cr
-1 & 1 } \biggr),
\end{eqnarray*}
where $\sigma_1^2, \sigma_2^2 > 1$.
When $\sigma_1^2 \neq\sigma_2^2$ and the sample size $n$ is large enough,
$X_1$ is $0$-stage second-order detectable (without conditioning on
any other predictor), and $X_2$ is $1$-stage second-order detectable
conditioning on $X_1$ because the conditional distribution, $X_2 |
X_1, Y \in S_h \sim\mathrm{N} ( (-1)^{h+1}X_1/\sigma_h^2,
1-1/\sigma_h^2 )$, is different for $h=1$ and $2$. Thus, both
$X_1$ and $X_2$ are stepwise detectable.
When $\sigma_1^2 =\sigma_2^2$, although $X_1$ and $X_2$ are relevant
predictors since the two conditional distributions are different, none
of them are
stepwise detectable. In this case, no stepwise procedure that selects
one variable at a time is able to ``detect'' either $X_1$ or $X_2$.
\end{example}

In Appendix~\ref{app5}, we prove the following theorem, which
guarantees that by appropriately choosing thresholds, the stepwise
procedure will keep adding predictors until all the stepwise detectable
predictors have been included, and keep removing predictors until all
the redundant variables have been excluded.

\begin{theorem}\label{thmsiri}Under model (\ref{eqaug}),
Conditions~\ref{cond1} and~\ref{cond2}, if all the relevant predictors
indexed by $\mathcal{A}$ are stepwise detectable with constant $\kappa
$, then there exists constant $c^*>0$ such that as $n \rightarrow\infty$,
\begin{eqnarray*}
& &\operatorname{Pr} \Bigl(\min_{\mathcal{C}\dvtx \mathcal{C}^c\cap\mathcal{A}
\neq\varnothing} \max
_{j \in\mathcal{C}^c} \widehat{D}^*_{j|\mathcal{C}} \geq c^*n^{-\kappa}
\Bigr)
\\
&&\qquad\geq 1-O \biggl(p(p+1) (H+1)\exp{ \biggl(-M_1
\frac
{n^{1-2\kappa}}{p^2H^2} \biggr)} \biggr) \rightarrow1
\end{eqnarray*}
and
\begin{eqnarray*}
& &\operatorname{Pr} \biggl(\max_{\mathcal{C}\dvtx \mathcal{C}^c\cap\mathcal{A} =
\varnothing} \max
_{j \in\mathcal{C}^c} \widehat{D}^*_{j|\mathcal{C}} < \frac{c^*}{2}n^{-\kappa}
\biggr)
\\
&&\qquad\geq 1-O \biggl(p(p+1) (H+1)\exp{ \biggl(-M_2
\frac
{n^{1-2\kappa}}{p^2H^2} \biggr)} \biggr) \rightarrow1,
\end{eqnarray*}
where $M_1$ and $M_2$ are positive constants.
\end{theorem}
 Therefore, by appropriately choosing the thresholds, the
stepwise procedure based on $\widehat{D}^*_{j|\mathcal{C}}$ is
consistent in identifying stepwise detectable predictors.

\subsection{Sure independence screening property of SIS$^*$}\label{secthr3}

%
\begin{definition}[(Individually detectable)]\label{defind}
We call a predictor $X_j$ individually detectable if there exist
constants $\xi_1,\xi_2 > 0$ and $\kappa\geq0$ such that either
%
\begin{equation}
\label{eqkappa2} \frac{\operatorname{Var} (\mathbb{E} (X_j | S(Y) ) )}{\mathbb
{E} (\operatorname{Var} (X_j | S(Y) ) )} \geq\xi_1 n^{-\kappa},
\end{equation}
or
\[
\log\mathbb{E} \bigl(\operatorname{Var} \bigl(X_j | S(Y) \bigr)
\bigr)-\mathbb {E}\log \bigl[\operatorname{Var} \bigl(X_j | S(Y)
\bigr) \bigr] \geq\xi_2 n^{-\kappa}.
\]
\end{definition}
 Simply put, individually detectable predictors have either
different means or different variances across slices. Therefore, in
the example (\ref{example}), both $X_1$ and $X_2$ are individually
detectable because $\operatorname{Var}(X_1|Y \in S_h)$ and $\operatorname{Var}(X_2|Y
\in S_h)$ ($1 \leq h \leq H$) are different across
slices. Note that not all stepwise detectable
predictors according to Definition~\ref{defstep} are individually
detectable. In Example~\ref{exex2} with $\sigma_1\neq\sigma_2$,
$X_2$ has the same distribution given $Y \in S_1$ or $Y \in S_2$,
but the conditional distributions of $X_2$ given $X_1$ are different
in two slices. That is, $X_2$ is stepwise detectable. However, an
independence screening method can only pick up variable $X_1$, but
not~$X_2$.

Theorem~\ref{thmsis}, which is proved in \citet{Jiang2014}, shows that
SIS$^*$ almost surely includes all the individually detectable
predictors under the following condition with ultra-high dimensionality
of predictors.
%
\begin{condition}\label{cond3}
$\log(p) = O(n^{\gamma})$ as $n \rightarrow\infty$ with $0 < \gamma+
2\kappa< 1$, where $\kappa$ is the same constant as in (\ref
{eqkappa2}). Furthermore, the number of the relevant predictors
$|\mathcal{A}| \leq n^{\eta}$ with $\eta+ 2\kappa< 1/2$.
\end{condition}

\begin{theorem}\label{thmsis}
Under Conditions~\ref{cond1} and~\ref{cond3}, if all the relevant
predictors indexed by $\mathcal{A}$ are individually detectable, then
there exist $c > 0$ and $C > 0$ such that
\begin{eqnarray*}
& &\operatorname{Pr} \Bigl(\min_{j \in\mathcal{A}} \widehat{D}^*_j
\geq c n^{-\kappa} \Bigr)
\\
&&\qquad\geq 1-O \biggl(p(H+1)\exp{ \biggl(-L_1\frac{n^{1-2\kappa
}}{H^2}
\biggr)} \biggr) \rightarrow1
\end{eqnarray*}
and
\begin{eqnarray*}
& &\operatorname{Pr} \bigl( \bigl\llvert \bigl\{ j\dvtx \widehat{D}^*_j
\geq cn^{-\kappa}, 1 \leq j \leq p \bigr\} \bigr\rrvert \leq C
n^{\kappa+\eta} \bigr)
\\
&&\qquad\geq 1-O \biggl(p(H+1)\exp{ \biggl(-L_2\frac{n^{1-2\kappa
}}{H^2}
\biggr)} \biggr) \rightarrow1,
\end{eqnarray*}
where $L_1$ and $L_2$ are positive constants.
\end{theorem}
 According to Theorem~\ref{thmsis}, we can first use SIS$^*$,
which is based on $\{\widehat{D}^*_j, 1\leq j \leq p\}$, to reduce the
dimensionality from $p$ to a scale between $n^{\kappa+\eta}$ and
$n^{{1}/{2}-\kappa}$ (since $\eta+\kappa<1/2-\kappa$ under
Condition~\ref{cond3}), and then apply the stepwise procedure proposed
in the previous sections, which is consistent with dimensionality below
$n^{{1}/{2}-\kappa}$. As discussed above, predictors that are
stepwise detectable according to Definition~\ref{defstep} are not
necessarily individually detectable. \citet{Fan2008} advocated an
iterative procedure that alternates between a large-scale screening and
a moderate-scale variable selection to enhance the performance, which
will be discussed in the next section.

\section{Implementation issues: Cross-stitching and
cross-validation}\label{secimp}

\begin{figure}

\includegraphics{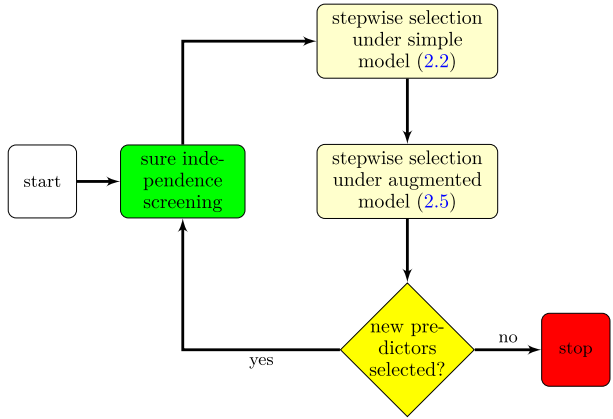}

\caption{Flowchart of SIRI.}\label{figsiri}
\end{figure}

The simple model (\ref{eqsim}) and the augmented model (\ref{eqaug})
compensate each other in terms of the bias-variance trade-off. Given
finite observations, model (\ref{eqsim}) is simpler and more powerful
when the response is driven by some linear combinations of covariates,
while model (\ref{eqaug}) is useful in detecting variables with more
complex relationships such as heteroscedastic effects or interactions.
Similarly, the SIS$^*$ procedure introduced in Section~\ref{secsis} is
very useful when we have a very large number of predictors, but it
cannot pick up stepwise detectable predictors that have the same
marginal distributions across slices. To find a balance between
simplicity and detectability, we propose the following cross-stitching strategy:
\begin{itemize}
\item \textit{Step} 0: initialize the current selection $\mathcal{C}=\varnothing$;
rank predictors according to $\{\widehat{D}^*_j, 1\leq j \leq p\}$ and
select a subset of predictors, denoted as $\mathcal{S}$, using SIS$^*$;
\item \textit{Step} 1: select predictors from set $\mathcal{S} \setminus\mathcal
{C}$ by using the stepwise procedure with addition and deletion steps
based on $\widehat{D}_{j|\mathcal{C}}$ in (\ref{eqhomo}) and add the
selected predictors into $\mathcal{C}$;
\item \textit{Step} 2: select predictors from set $\mathcal{S} \setminus\mathcal
{C}$ by using the stepwise procedure with addition and deletion steps
based on $\widehat{D}^*_{j|\mathcal{C}}$ in (\ref{eqheter}) and add
the selected predictors into $\mathcal{C}$;
\item \textit{Step} 3: conditioning on the current selection $\mathcal{C}$, rank
the remaining predictors based on $\{\widehat{D}^*_{j|\mathcal{C}}, j
\notin\mathcal{C}\}$, update set $\mathcal{S}$ using SIS$^*$, and
iterate steps 1--3 until no more predictors are selected.
\end{itemize}
We name the proposed procedure sliced inverse regression for
variable selection via inverse modeling, or SIRI for short. A flowchart
of the SIRI procedure is illustrated in Figure~\ref{figsiri}.

%

Theoretically, step $2$ is able to detect both linear and more complex
relationships and $\widehat{D}^*_{j|\mathcal{C}}$ picks up a larger set
than $\widehat{D}_{j|\mathcal{C}}$ does. However, in practice, we have
to use a relatively large threshold in step 2 to control the number of
false positives and thus may falsely discard linear predictors when
their effects are weak. Empirically, we have found that adding step 1
will enhance the performance of SIRI in linear or near-linear models,
while having almost no effects on its performances in complex models
with interaction or other second-order terms.

In the addition step of the stepwise procedure, instead of selecting
the variable from $\mathcal{C}^c$ with the maximum value of $\widehat
{D}_{j|\mathcal{C}}$ (or $\widehat{D}^*_{j|\mathcal{C}}$), we may also
sequentially add variables with $\widehat{D}_{j|\mathcal{C}} > \nu_a$
(or $\widehat{D}^*_{j|\mathcal{C}} > \nu_a^*$). Specifically, given
thresholds $\nu_a>\nu_d$ and the current set of selected predictors
indexed by $\mathcal{C}$, we can modify each iteration of the original
stepwise procedure as following:
\begin{itemize}
\item Modified addition step: for each variable $j \in\{1,\ldots,p\}$,
let $\mathcal{C}=\mathcal{C}+\{j\}$ if $j \notin\mathcal{C}$ and
$\widehat{D}_{j|\mathcal{C}} > \nu_a$.
\item Deletion step: find $j_d$ such that $\widehat{D}_{j_d|\mathcal
{C}-\{j_d\}} = \min_{j \in\mathcal{C}} \widehat{D}_{j|\mathcal{C}-\{j\}
}$; let $\mathcal{C} = \mathcal{C}-\{j_d\}$ if $\widehat
{D}_{j_d|\mathcal{C}-\{j_d\}} < \nu_d$.
\end{itemize}
The stepwise procedure with the modified addition step may use fewer
iterations to find all relevant predictors and will not stop until all
relevant predictors have been included if we choose $\nu_a = c
n^{-\kappa}$ in Theorem~\ref{thmsim}. However, in practice, the
performance of the modified procedure depends on the ordering of the
variables and is less stable than the original procedure. Since we are
less concerned about the computational cost of SIRI, we implement the
original addition step in the following study.

In our previous discussions, we have assumed that a fixed slicing
scheme is given. In practice, we need to choose a slicing scheme. If we
assume that there is a true slicing scheme from which data are
generated, \citet{Jiang2014} showed that the power of the stepwise
procedure tends to increase with a larger number of slices, but there
is no gain by further increasing the number of slices once the slicing
is already more refined than the true slicing scheme. In practice, the
true slicing scheme is usually unknown (except maybe in cases when the
response is discrete). When a slicing scheme uses a larger number of
slices, the number of observations in each slice decreases, which makes
the estimation of parameters in the model less accurate and less
stable. We observed from intensive simulation studies that, with a
reasonable number of observations in each slice (say 40 or more), a
larger number of slices is preferred.

We also need to choose the number of effective directions $q$ in model
(\ref{eqsim}) and thresholds for deciding to add or to delete
variables. Sections~\ref{secmodel} and~\ref{sectheory} characterize
asymptotic distributions and behaviors of stepwise procedures, and
provide some theoretical guidelines for choosing the thresholds.
However, these theoretical results are not directly usable because: (1)
the asymptotic distributions that we derived in (\ref{eqdist1}) and
(\ref{eqdist2}) are for a single addition or deletion step; (2) the
consistency results are valid in asymptotic sense and the rate of
increase in dimension relative to sample size is usually unknown. In
practice, we propose to use a $K$-fold cross-validation (CV) procedure
for selecting thresholds and the number of effective directions $q$.

We consider two performance measures for $K$-fold cross-validations:
classification error (CE) and mean absolute error (AE). Suppose there
are $n$ training samples and $m$ testing samples. The $j$th observation
($j=1,2,\ldots,m$) in the testing set has response $y_j$ and slice
membership $S(y_j)$ (the slicing scheme is fixed based on training
samples). Let $p_j^{(h)} = \operatorname{Pr} (S(y_j)=h|\mathbf{X}=\bx
_j,\widehat{\theta} )$ be the estimated probability that the
observation $j$ is from slice $S_h$, where $\widehat{\theta}$ denotes
the maximum likelihood estimate of model parameters. The classification
error is defined as
\[
\operatorname{CE} = \frac{1}{m}\sum_{j=1}^m
\mathbb{I} \Bigl[S(y_j) \neq \mathop{\operatorname{argmax}}_{h}
\bigl(p_j^{(h)} \bigr) \Bigr].
\]
We denote the average response of training samples in slice $S_h$ as
\[
\bar{y}^{(h)}=
\frac{\sum_{i=1}^n\mathbb{I} [S(y_i)=h ]y_i}{\sum_{i=1}^n\mathbb{I}
[S(y_i)=h ]},\qquad h=1,2,\ldots,H.
\]
The absolute error is defined as
\[
\operatorname{AE} = \frac{1}{m}\sum_{j=1}^m
\Biggl\llvert y_j - \sum_{h=1}^H
p_j^{(h)}\bar{y}^{(h)} \Biggr\rrvert.
\]
CE is a more relevant performance measure when the response is
categorical or there is a nonsmooth functional relationship (e.g.,
rational functions) between the response and predictors, and AE is a
better measure when there is a monotonic and smooth functional
relationship between the response and predictors. There are other
measures that have compromising features between these two measures,
such as median absolute deviation, which will not be explored here. We
will use CE and AE as performance measures throughout simulation
studies and name the corresponding methods SIRI-AE and SIRI-CE, respectively.

\section{Simulation studies}\label{secsimulation}

In order to facilitate fair comparisons with other existing methods
that are motivated from the forward modeling perspective, examples
presented here are all generated under forward models, which violates
the basic model assumption of SIRI. The setting of the simulation also
demonstrates the robustness of SIRI when some of its model assumptions
are violated, especially the normality assumption on relevant predictor
variables within each slice.


\subsection{Independence screening performance}\label{secscreen}
We first compare the variable screening performance of SIRI with
iterative sure independence screening (ISIS) based on correlation
learning proposed by \citet{Fan2008} and sure independence screening
based on distance correlation (DC-SIS) proposed by \citet{Li2012}. We
evaluate the performance of each method according to the proportion
that relevant predictors are placed among the top $[n/\log(n)]$
predictors ranked by it, with larger values indicating better
performance in variable screening.

In the simulation, the predictor variables $\mathbf{ X}=(X_1,X_2,\ldots,X_p)^T$ were generated from a $p$-variate normal distribution with
meanu $0$ and covariances $\operatorname{Cov} (X_i,X_j )=\rho^{|i-j|}$
for $1\leq i,j \leq p$. We generate the response variable from the
following three scenarios:
\begin{eqnarray*}
&&\mbox{Scenario } 0.1\mbox{:}\qquad  Y=X_2-\rho
X_1+0.2X_{100}+\sigma\varepsilon,
\\
&&\mbox{Scenario } 0.2\mbox{:}\qquad  Y=X_1X_2+\sigma
e^{2|X_{100}|} \varepsilon,
\\
&&\mbox{Scenario } 0.3\mbox{:}\qquad Y=\frac{X_{100}}{X_1+X_2}+\sigma\varepsilon,
\end{eqnarray*}
where sample size $n=200$, $\sigma=0.2$, and $\varepsilon\sim N(0,1)$
independent of $\mathbf{X}$. For each scenario, we simulated $100$ data
sets according to six different settings with dimension $p=2000$ or
$5000$ and correlation $\rho=0.0$, $0.5$ or $0.9$. Scenario $0.1$ is a
linear model with three additive effects. The way $X_1$ is introduced
is to make it marginally uncorrelated with the response $Y$ (note that
when $\rho=0.0$, $X_1$ is not a relevant predictor). We added another
variable $X_{100}$ that has negligible correlation with $X_1$ and $X_2$
and a very small correlation with the response $Y$. Scenario $0.2$
contains an interaction term $X_1X_2$ and a heteroscedatic noise term
determined by~$X_{100}$. Scenario $0.3$ is an example of a rational
model with interactions.

\begin{table}
\caption{The proportions that relevant predictors are placed among the
top $[n/\log(n)]$ by different screening methods under Scenarios
0.1--0.3 in Section~\protect\ref{secscreen}}
\label{tabscreen}
\begin{tabular*}{\textwidth}{@{\extracolsep{\fill}}lccccccccc@{}}
%
\hline
 & \multicolumn{3}{c}{\textbf{Scenario 0.1}} & \multicolumn{3}{c}{\textbf{Scenario 0.2}} &
\multicolumn{3}{c}{\textbf{Scenario 0.3}} \\[-6pt]
& \multicolumn{3}{c}{\hrulefill} & \multicolumn{3}{c}{\hrulefill} &
\multicolumn{3}{c@{}}{\hrulefill} \\
\textbf{Method}& $\bolds{X_1}$ & $\bolds{X_2}$ & $\bolds{X_{100}}$ & $\bolds{X_1}$ & $\bolds{X_2}$ & $\bolds{X_{100}}$ & $\bolds{X_1}$ &
$\bolds{X_2}$ & $\bolds{X_{100}}$ \\
\hline
\multicolumn{10}{c}{{Setting 1}: $p=2000$, $\rho=0.0$} \\
ISIS & -- & 1.00 & 1.00 & 0.02 & 0.01 & 0.46 & 0.00 & 0.00 & 0.09 \\
DC-SIS & -- & 1.00 & 0.55 & 0.07 & 0.09 & 1.00 & 0.00 & 0.00 & 0.60 \\
SIRI & -- & 1.00 & 0.30 & 0.32 & 0.25 & 0.97 & 1.00 & 0.99 & 1.00 \\[3pt]
\multicolumn{10}{c}{{Setting 2}: $p=2000$, $\rho=0.5$} \\
ISIS & 1.00 & 1.00 & 1.00 & 0.04 & 0.02 & 0.54 & 0.00 & 0.00 & 0.15 \\
DC-SIS & 0.02 & 1.00 & 0.71 & 0.55 & 0.53 & 1.00 & 0.03 & 0.00 & 0.59
\\
SIRI & 1.00 & 1.00 & 0.45 & 0.92 & 0.87 & 0.92 & 1.00 & 1.00 & 1.00 \\[3pt]
\multicolumn{10}{c}{{Setting 3}: $p=2000$, $\rho=0.9$} \\
ISIS & 0.93 & 0.98 & 0.91 & 0.03 & 0.02 & 0.55 & 0.00 & 0.00 & 0.04 \\
DC-SIS & 0.01 & 0.99 & 1.00 & 0.96 & 0.95 & 1.00 & 0.34 & 0.38 & 0.63
\\
SIRI & 0.93 & 0.82 & 0.79 & 0.99 & 0.56 & 0.95 & 0.98 & 0.98 & 1.00 \\[3pt]
\multicolumn{10}{c}{{Setting 4}: $p=5000$, $\rho=0.0$} \\
ISIS & -- & 1.00 & 1.00 & 0.02 & 0.00 & 0.43 & 0.00 & 0.00 & 0.06 \\
DC-SIS & -- & 1.00 & 0.39 & 0.03 & 0.05 & 1.00 & 0.00 & 0.00 & 0.44 \\
SIRI & -- & 1.00 & 0.14 & 0.15 & 0.16 & 0.99 & 0.99 & 1.00 & 1.00 \\[3pt]
\multicolumn{10}{c}{{Setting 5}: $p=5000$, $\rho=0.5$} \\
ISIS & 1.00 & 1.00 & 1.00 & 0.03 & 0.02 & 0.60 & 0.00 & 0.00 & 0.07 \\
DC-SIS & 0.05 & 1.00 & 0.71 & 0.41 & 0.44 & 1.00 & 0.00 & 0.02 & 0.61
\\
SIRI & 1.00 & 1.00 & 0.39 & 0.88 & 0.86 & 0.94 & 0.98 & 1.00 & 0.99 \\[3pt]
\multicolumn{10}{c}{{Setting 6}: $p=5000$, $\rho=0.9$} \\
ISIS & 0.86 & 0.99 & 0.87 & 0.02 & 0.03 & 0.34 & 0.00 & 0.00 & 0.03 \\
DC-SIS & 0.01 & 0.99 & 0.99 & 0.92 & 0.93 & 1.00 & 0.22 & 0.13 & 0.49
\\
SIRI & 0.82 & 0.79 & 0.74 & 0.95 & 0.53 & 0.90 & 0.85 & 0.99 & 1.00 \\
\hline
\end{tabular*}
%
\end{table}

Proportions that relevant predictors are placed among the top $[n/\log
(n)]$ by different screening methods are shown in Table~\ref{tabscreen}. Under Scenario $0.1$ with linear models, we can see that
ISIS and DC-SIS had better power than SIRI in detecting variables that
are weakly correlated with the response ($X_{100}$ in this example).
When predictors are correlated (Settings 2--3 and 4--5), iterative
procedures, ISIS and SIRI, were more effective in detecting variables
that are marginally uncorrelated with the response ($X_1$ in this
example) compared with DC-SIS. Under Scenario $0.2$, ISIS based on
linear models failed to detect the variables in the interaction term
and often misses the predictor in the heteroscedastic noise term. When
there are moderate correlations between two predictors $X_1$ and $X_2$
in the interaction term (Settings $2$ and $4$), DC-SIS picked up $X_1$
and $X_2$ about half of the time. However, when the two predictors are
uncorrelated (Settings $1$ and $3$), DC-SIS failed to detect them. SIRI
outperformed DC-SIS in detecting variables with interactions for both
settings with $\rho=0.0$ and $\rho=0.5$. Note that when there is a
strong correlation between two predictors, say $X_1$ and $X_2$ (Settings
$3$ and $5$), each model can be approximated well by a reduced model
under the constraint $X_2=c X_1$. In this case, the noniterative
procedure DC-SIS is able to pick up both variables, but SIRI sometimes
missed one of the variables since it treats the other variable as
redundant, which perhaps is the correct decision. We also notice that
the noniterative version of SIRI is able to detect both $X_1$ and $X_2$
more often than DC-SIS (results not shown here). Under Scenario $0.3$,
when there is a rational relationship between the response and the
relevant predictors, SIRI significantly outperformed the other two
methods in detecting the relevant predictors. Performances of different
methods are only slightly affected as we increase the dimension from
$p=2000$ to $p=5000$.

\subsection{Variable selection performance}\label{secselect}

We further study the variable selection accuracy of SIRI and other
existing methods in identifying relevant predictors and excluding
irrelevant predictors. In the following examples, for both SIRI and
COP, we implemented a fixed slicing scheme with $5$ slices of equal
size ({i.e.}, $H=5$) and used a $10$-fold CV procedure to
determine the stepwise variable selection thresholds and the number of
effective directions $q$ in model (\ref{eqsim}) of Section~\ref{seclrt}. Specifically, the number of effective directions $q$ was
chosen from $\{0,1,2,3,4\}$, where $q=0$ means that we skipped the
variable selection step under simple model~(\ref{eqsim}) in the
iterative procedure described by Figure~\ref{figsiri}. The thresholds
in addition and deletion steps were selected from the grid $\{(\nu
_{i,a} = \chi^2(\alpha_i,q), \nu_{i,d} = \chi^2(\alpha_i-0.05,q))\}$
for simple model (\ref{eqsim}) and from the grid $\{(\nu^*_{i,a} =
\frac{n}{n-H(d+2)}\chi^2(\alpha_i,(H-1)(d+2)), \nu^*_{i,d} = \frac
{n}{n-H(d+2)}\chi^2(\alpha_i-0.05,(H-1)(d+2)))\}$ for augmented model~(\ref{eqaug}), where $\chi^2(\alpha,\mathrm{d.f.})$ is the $100\alpha$th
quantile of $\chi^2(\mathrm{d.f.})$ and $d=|\mathcal{C}|$ is the number
of previously selected predictors. For a given $p$, the dimension of
predictors, we chose $\{\alpha_i\} = \{
1-p^{-1},1-0.5p^{-1},1-0.1p^{-1},1-0.05p^{-1},1-0.01p^{-1}\}$.

The other variable selection methods to be compared with SIRI and COP
include Lasso, ISIS-SCAD (SCAD with iterative sure independence
screening), SpAM and hierNet, which is a Lasso-like procedure to detect
multiplicative interactions between predictors under hierarchical
constraints. The R packages glmnet, SIS, COP, SAM and hierNet are used
to run Lasso, ISIS-SCAD, COP, SpAM and hierNet, respectively. For Lasso
and hierNet, we select the largest regularization parameter with
estimated CV error less than or equal to the minimum estimated CV error
plus one standard deviation of the estimate. The tuning parameters SCAD
and SpAM are also selected by CV.

For variable selections under index models with linear or first-order
effects, we generated the predictor variables $\mathbf{ X}=(X_1,X_2,\ldots,X_p)^T$ from a multivariate normal distribution with mean $0$ and
covariances $\operatorname{Cov} (X_i,X_j )=\rho^{|i-j|}$ for $1\leq
i,j \leq p$, and simulated the response variable according to the
following models:
\begin{eqnarray*}
&&\mbox{Scenario } 1.1\mbox{:}\qquad  Y=\bolds\beta^T\mathbf{X}+
\sigma\varepsilon,\qquad n=200, \sigma=1.0, \rho=0.5,
\\
& &\hspace*{84pt} \bolds\beta=(3,1.5,2,2,2,2,2,2,0,\ldots,0),
\\
&&\mbox{Scenario } 1.2\mbox{:} \qquad Y=\frac{\sum_{j=1}^3X_j}{0.5+(1.5+\sum_{j=2}^4X_j)^2}+\sigma\varepsilon,
\\
& &\hspace*{84pt} n=200, \sigma=0.2, \rho=0.0,
\\
&&\mbox{Scenario } 1.3\mbox{:} \qquad Y=\frac{\sigma\varepsilon}{1.5+\sum_{j=1}^8X_j},\qquad n=1000, \sigma=0.2, \rho=0.0,
\end{eqnarray*}
where $n$ is the number of observations, $p$ is the number of
predictors and is set as $1000$ here, and the noise $\varepsilon$ is
independent of $\mathbf{X}$ and follows $N(0,1)$. Scenario $1.1$ is a
linear model which involves $8$ true predictors and $992$ irrelevant
predictors. Scenario $1.2$, a multi-index model with $4$ true
predictors, was studied in \citet{Li1991} and \citet{Zhong2012}, and
there is a nonlinear relationship between the response $Y$ and two
linear combinations of predictors $X_1+X_2+X_3$ and $X_2+X_3+X_4$.
Scenario $1.3$ is a single-index model with $8$ true predictors and
heteroscedastic noise.

For each simulation setting, we randomly generated $100$ data sets each
with $n$ observations and applied variable selection methods to each
data set. Two quantities, the average number of irrelevant predictors
falsely selected as true predictors (which is referred to as FP) and
the average number of true predictors falsely excluded as irrelevant
predictors (which is referred to as FN), were used to measure the
variable selection performance of each method. For example, under
Scenario~$1.1$, the FPs and FNs range from 0 to 992 and from 0 to 8,
respectively, with smaller values indicating better accuracies in
variable selection. The FP- and FN-values of different methods together
with their corresponding standard errors (in brackets) are reported in
Table~\ref{tabindex}.

\begin{table}
\tabcolsep=0pt
\caption{False positive (FP) and false negative (FN) values of
different variable selection methods under Scenarios 1.1--1.3}
\label{tabindex}
\begin{tabular*}{\textwidth}{@{\extracolsep{\fill}}lcccccc@{}}
\hline
& \multicolumn{2}{c}{\textbf{Scenario 1.1}} & \multicolumn
{2}{c}{\textbf{Scenario 1.2}} & \multicolumn{2}{c@{}}{\textbf{Scenario 1.3}} \\[-6pt]
 & \multicolumn{2}{c}{\hrulefill} & \multicolumn
{2}{c}{\hrulefill} & \multicolumn{2}{c@{}}{\hrulefill} \\
\textbf{Method}
& \multicolumn{1}{c}{\textbf{FP} $\bolds{(0, 992)}$} & \multicolumn{1}{c}{\textbf{FN} $\bolds{(0, 8)}$} & \multicolumn{1}{c}{\textbf{FP} $\bolds{(0, 996)}$} &
\multicolumn{1}{c}{\textbf{FN} $\bolds{(0, 4)}$} & \multicolumn{1}{c}{\textbf{FP} $\bolds{(0, 992)}$} & \multicolumn{1}{c@{}}{\textbf{FN} $\bolds{(0, 8)}$} \\
\hline
Lasso & 0.59 (0.10) & 0.00 (0.00) & 0.08 (0.03) & 1.07 (0.03) & 0.00
(0.00) & 8.00 (0.00) \\
ISIS-SCAD & 0.35 (0.07) & 0.00 (0.00) & 0.60 (0.08) & 1.02 (0.01) &
5.08 (0.65) & 7.97 (0.02) \\
hierNet & 1.49 (0.19) & 0.00 (0.00) & 8.72 (0.36) & 0.93 (0.03) & 7.68
(0.48) & 7.94 (0.02) \\
SpAM & 1.29 (0.19) & 0.00 (0.00) & 2.44 (0.20) & 0.84 (0.04) & 2.49
(0.16) & 7.99 (0.01) \\
COP & 0.69 (0.12) & 0.06 (0.03) & 1.84 (0.16) & 0.98 (0.01) & 1.26
(0.13) & 3.32 (0.19) \\
SIRI-AE & 0.01 (0.01) & 0.09 (0.04) & 0.13 (0.04) & 0.07 (0.03) & 0.43
(0.08) & 4.82 (0.27) \\
SIRI-CE & 0.26 (0.05) & 0.08 (0.03) & 0.55 (0.08) & 0.09 (0.03) & 2.02
(0.17) & 0.51 (0.16) \\
\hline
\end{tabular*}
%
\end{table}

Under Scenario $1.1$, variable selection methods derived from additive
models (Lasso, SCAD, SpAM and hierNet) were able to detect all the
relevant predictors (FN${}={}$0) with few false positives. On the other hand,
COP, SIRI-AE and SIRI-CE missed some (about $10\%$) relevant predictors
while excluded most irrelevant ones (lower FP values). The relatively
high accuracy of methods developed for linear models is expected under
this scenario, because the observations were simulated from a linear
relationship. Under Scenario $1.2$, Lasso achieved the lowest false
positives, but it almost always missed one of the relevant predictor,
$X_4$, because of its nonlinear relationship with the response.\vadjust{\goodbreak} The
other methods developed under the linear model assumption suffered from
the same issue. However, SIRI-AE and SIRI-CE was able to detect most of
the four relevant predictors (FN${}={}$0.09 and 0.07) with a comparable
number of false positives. Under the heteroscedastic model in Scenario
$1.3$, the methods based on linear models failed to detect relevant
predictors. Among other methods, SIRI-AE achieved the lowest number of
false positives (FP${}={}$0.43) but missed about half of the relevant
predictors (FN${}={}$4.82), while SIRI-CE selected most of the relevant
predictors (FN${}={}$0.51) with a reasonably low false positives (FP${}={}$2.02).
The performance of COP was in-between SIRI-AE and SIRI-CE with FN${}={}$3.32
and FP${}={}$1.26. A possible explanation for the better performance of
SIRI-CE relative to SIRI-AE in this setting is because the generative
model under Scenario $1.3$ contains a singular point at $\sum_{j=1}^8X_j=-1.5$.
Since the absolute error is less robust to outliers
than the classification error, SIRI-AE is more sensitive to the
inclusion of irrelevant predictors and more conservative in selecting
predictors.

Next, we consider forward models containing variables with higher-order
effects. Predictor variables $X_1,X_2,\ldots,X_p$ were independent and
identically distributed $N(0,1)$ random variables, and the response was
generated under the following models given the predictors:
\begin{eqnarray*}
&&\mbox{Scenario } 2.1\mbox{:}\qquad Y=\alpha X_1 + \alpha
X_2 + X_1X_2+\sigma \varepsilon, \qquad \alpha=0.2, n=200,
\\
&&\mbox{Scenario } 2.2\mbox{:} \qquad Y=X_1+X_1X_2+X_1X_3+
\sigma\varepsilon,\qquad n=200,
\\
&&\mbox{Scenario } 2.3\mbox{:} \qquad Y=X_1X_2+X_1X_3+
\sigma\varepsilon,\qquad n=200,
\\
&&\mbox{Scenario } 2.4\mbox{:} \qquad Y=X_1X_2X_3+
\sigma\varepsilon,\qquad n=200, 500\mbox{ and }1000,
\\
&&\mbox{Scenario } 2.5\mbox{:} \qquad Y=X_1^2X_2+
\sigma\varepsilon,\qquad n=200,
\\
&&\mbox{Scenario } 2.6\mbox{:} \qquad Y=\frac{X_1}{X_2+X_3}+\sigma\varepsilon,\qquad
n=200,
\end{eqnarray*}
where $n$ is the number of observations, $p$ is the number of
predictors and is set as $1000$ here, $\sigma=0.2$ and $\varepsilon$ is
independent of $\mathbf{X}$ and follows $N(0,1)$. The models under
Scenarios $2.1$ and $2.2$ have strong (both and $X_1$ and $X_2$ have
main effects in Scenario $2.1$) and weak (only $X_1$ has main effect in
Scenario $2.2$) hierarchical interaction terms, respectively. Scenario
$2.3$ contains predictors with pairwise multiplicative interactions and
without main effects. The three-way interaction model in Scenario $2.4$
was simulated under three settings with different sample sizes:
$n=200$, $n=500$ and $n=1000$. Scenario $2.5$ contains a quadratic
interaction term and Scenario $2.6$ has a rational relationship.

\begin{table}
\tabcolsep=0pt
\caption{False positive (FP) and false negative (FN) values of
different variable
selection methods under Scenarios 2.1--2.3}
\label{tabinteract1}
\begin{tabular*}{\textwidth}{@{\extracolsep{\fill}}lcccccc@{}}
\hline
& \multicolumn{2}{c}{\textbf{Scenario 2.1}} & \multicolumn
{2}{c}{\textbf{Scenario 2.2}} & \multicolumn{2}{c@{}}{\textbf{Scenario 2.3}} \\[-6pt]
& \multicolumn{2}{c}{\hrulefill} & \multicolumn
{2}{c}{\hrulefill} & \multicolumn{2}{c@{}}{\hrulefill} \\
\textbf{Method}
& \multicolumn{1}{c}{\textbf{FP} $\bolds{(0, 998)}$} & \multicolumn{1}{c}{\textbf{FN} $\bolds{(0, 2)}$} & \multicolumn{1}{c}{\textbf{FP} $\bolds{(0, 997)}$} &
\multicolumn{1}{c}{\textbf{FN} $\bolds{(0, 3)}$} & \multicolumn{1}{c}{\textbf{FP} $\bolds{(0, 997)}$} & \multicolumn{1}{c@{}}{\textbf{FN} $\bolds{(0, 3)}$} \\
\hline
ISIS-SCAD-$2$ & 0.00 (0.00) & 0.00 (0.00) & 0.00 (0.00) & 0.06 (0.04)
& 0.00 (0.00) & 0.03 (0.03) \\
DC-SIS-SCAD-$2$ & 0.00 (0.00) & 0.00 (0.00) & 0.25 (0.09) & 0.11
(0.03) & 1.56 (0.19) & 1.81 (0.11) \\
hierNet & 10.45 (0.57)\phantom{0} & 0.00 (0.00) & 10.34 (0.71)\phantom{0} &
0.02 (0.05) &
12.17 (0.73)\phantom{0} & 0.04 (0.03) \\
SpAM & 2.35 (0.30) & 1.18 (0.05) & 0.03 (0.02) & 1.99 (0.01) & 4.44
(0.29) & 2.66 (0.05) \\
SIRI-AE & 0.00 (0.00) & 0.00 (0.00) & 0.02 (0.01) & 0.04 (0.02) & 0.10
(0.04) & 0.11 (0.05) \\
SIRI-CE & 0.64 (0.11) & 0.00 (0.00) & 0.29 (0.06) & 0.10 (0.04) & 0.86
(0.12) & 0.11 (0.05) \\
\hline
\end{tabular*}
%
\end{table}

\begin{table}[b]
\tabcolsep=0pt
\caption{False positive (FP) and false negative (FN) values of
different variable selection methods under Scenario $2.4$ with
different sample sizes}
\label{tabinteract2}
\begin{tabular*}{\textwidth}{@{\extracolsep{4in minus 4in}}lcccccc@{}}
\hline
 & \multicolumn{2}{c}{\textbf{Scenario 2.4 (}$\bolds{n=200}$\textbf{)}} & \multicolumn
{2}{c}{\textbf{Scenario 2.4 (}$\bolds{n=500}$\textbf{)}} & \multicolumn{2}{c@{}}{\textbf{Scenario 2.4 (}$\bolds{n=1000}$\textbf{)}} \\[-6pt]
 & \multicolumn{2}{c}{\hrulefill} & \multicolumn
{2}{c}{\hrulefill} & \multicolumn{2}{c@{}}{\hrulefill} \\
\multicolumn{1}{@{}l}{\textbf{Method}} & \multicolumn{1}{c}{\textbf{FP} $\bolds{(0, 997)}$} &
\multicolumn{1}{c}{\textbf{FN} $\bolds{(0, 3)}$} &
\multicolumn{1}{c}{\textbf{FP} $\bolds{(0, 997)}$} &
\multicolumn{1}{c}{\textbf{FN} $\bolds{(0, 3)}$} &
\multicolumn{1}{c}{\textbf{FP} $\bolds{(0,997)}$} &
\multicolumn{1}{c@{}}{\textbf{FN} $\bolds{(0, 3)}$} \\
\hline
DC-SIS-SCAD-$3$ & 0.45 (0.12) & 0.85 (0.12) & 0.00 (0.00) & 0.00
(0.00) & 0.00 (0.00) & 0.00 (0.00) \\
hierNet & 7.99 (0.65) & 2.29 (0.08) & 7.83 (1.17) & 2.37 (0.08) & 3.66
(1.09) & 2.61 (0.06) \\
SpAM & 3.40 (0.27) & 2.54 (0.06) & 3.22 (0.30) & 2.43 (0.07) & 4.19
(0.42) & 2.32 (0.07) \\
SIRI-AE & 0.98 (0.12) & 2.27 (0.06) & 0.36 (0.09) & 0.70 (0.07) & 0.21
(0.06) & 0.00 (0.00) \\
SIRI-CE & 1.98 (0.16) & 2.27 (0.07) & 1.96 (0.17) & 0.46 (0.05) & 2.03
(0.19) & 0.00 (0.00) \\
\hline
\end{tabular*}
%
\end{table}

Because methods such as Lasso and SCAD are not specifically designed
for detecting variables with nonlinear effects and are clearly at a
disadvantage, we did not directly compare them with SIRI, SpAM and
hierNet. For the purpose of comparison, we created a benchmark method
based on ISIS-SCAD by applying ISIS-SCAD to an expanded set of
predictors that includes all the terms up to $k$-way multiplicative
interactions. The corresponding method, which we referred to as
ISIS-SCAD-$k$, is an oracle benchmark under Scenarios $2.1$--$2.4$
where responses were generated according to $2$-way or $3$-way
multiplicative interactions. Since DC-SIS as a screening tool has the
ability to detect individual predictors under the presence of
second-order effects, we also augmented ISIS-SCAD with DC-SIS and
denoted the method as DC-SIS-SCAD-$k$. In DC-SIS-SCAD-$k$, we first
used DC-SIS to reduce the number of predictors. Then we expanded the
selected predictors to include up to $k$-way multiplicative
interactions among them and applied ISIS-SCAD. Because DC-SIS-SCAD-$k$
does not need to consider all the interaction terms among $p$
predictors, it has a huge speed advantage over ISIS-SCAD-$k$ but it may
fail to detect all the predictors if the DC-SIS step does not retain
all the relevant predictors. The FP- and FN-values (and their standard
errors) of different methods including ISIS-SCAD-$k$ and
DC-SIS-SCAD-$k$ under various scenarios are shown in Tables~\ref{tabinteract1},
\ref{tabinteract2} and~\ref{tabinteract3}, respectively. Note that FP- and FN-values are
calculated based on the number of predictors selected by a method, not
based on the number of parameters used in building the model. For
example, if $X_3$, $X_4$ and $X_3X_4$ all have nonzero coefficients
from hierNet under Scenario~$2.1$, we count the number of false
positives as $2$, not $3$. Under Scenarios $2.1$--$2.3$, we also
compared the performances of SIRI-AE, SIRI-CE and DC-SIS-SCAD-$2$ when
the predictors are correlated [see Table~$9$ of \citet{Jiang2014}]. In
addition, to investigate the performance of SIRI with nonnormally
distributed predictor, we simulated Scenarios $2.1$--$2.3$ by
generating predictors from the uniform distribution on $(-2,2)$, and
the results are reported in Table~$9$ of \citet{Jiang2014}.

\begin{table}
\caption{False positive (FP) and false negative (FN) values of
different variable selection methods Scenarios~$2.5$ and $2.6$}
\label{tabinteract3}
\begin{tabular*}{\textwidth}{@{\extracolsep{\fill}}lcccc@{}}
\hline
 & \multicolumn{2}{c}{\textbf{Scenario 2.5}} & \multicolumn
{2}{c@{}}{\textbf{Scenario 2.6}} \\[-6pt]
 & \multicolumn{2}{c}{\hrulefill} & \multicolumn
{2}{c@{}}{\hrulefill} \\
\textbf{Method} & \multicolumn{1}{c}{\textbf{FP} $\bolds{(0, 998)}$} & \multicolumn{1}{c}{\textbf{FN} $\bolds{(0, 2)}$} &
\multicolumn{1}{c}{\textbf{FP} $\bolds{(0, 997)}$} & \multicolumn{1}{c@{}}{\textbf{FN} $\bolds{(0, 3)}$} \\
\hline
ISIS-SCAD-$2$ & 0.04 (0.02) & 1.09 (0.04) & 0.00 (0.00) & 3.00 (0.00)
\\
DC-SIS-SCAD-$2$ &2.38 (0.18) & 0.51 (0.05) & 0.81 (0.16) & 2.96 (0.02)
\\
hierNet & 0.06 (0.03) & 0.97 (0.02) & 6.18 (0.68) & 2.92 (0.03) \\
SpAM & 0.42 (0.09) & 0.83 (0.04) & 4.56 (0.32) & 1.58 (0.06) \\
SIRI-AE & 0.08 (0.03) & 0.00 (0.00) & 0.51 (0.11) & 0.00 (0.00) \\
SIRI-CE & 0.88 (0.11) & 0.01 (0.01) & 0.56 (0.11) & 0.00 (0.00) \\
\hline
\end{tabular*}
\end{table}

\begin{table}[b]
\caption{Average running time (in seconds) of different variable
selection methods under Scenarios $2.1$--$2.3$, $2.5$ and $2.6$}
\label{tabtime}
\begin{tabular*}{\textwidth}{@{\extracolsep{\fill}}ld{5.2}d{5.2}d{5.2}d{5.2}d{5.2}@{}}
\hline
\textbf{Method} & \multicolumn{1}{c}{\textbf{Scenario 2.1}} & \multicolumn{1}{c}{\textbf{Scenario 2.2}} &
 \multicolumn{1}{c}{\textbf{Scenario 2.3}} & \multicolumn{1}{c}{\textbf{Scenario 2.5}} &
\multicolumn{1}{c@{}}{\textbf{Scenario 2.6}} \\
\hline
ISIS-SCAD-$2$ & 14\mbox{,}279.11 & 9406.27 & 11\mbox{,}581.55 & 10\mbox{,}232.31 & 4220.24 \\
DC-SIS-SCAD-$2$ & 29.47 & 25.77 & 31.90 & 37.03 & 25.68 \\
hierNet & 16\mbox{,}625.38 & 26\mbox{,}171.28 & 34\mbox{,}733.13 & 37\mbox{,}312.59 & 27\mbox{,}255.16\\
SpAM & 5.91 & 4.57 & 5.40 & 4.72 & 4.65 \\
SIRI & 28.86 & 44.85 & 20.01 & 44.36 & 35.26\\
\hline
\end{tabular*}
%
\end{table}

Under Scenarios 2.1--2.3 of Table~\ref{tabinteract1}, the oracle
benchmark, ISIS-SCAD-2, correctly discovered most of the relevant
predictors with two-way interactions and did not pick up any irrelevant
predictor. It is encouraging to see that the performance of the
proposed method SIRI-AE was comparable with ISIS-SCAD-2 (in terms of
both false positives and false negatives), although SIRI-AE did not
assume the knowledge on the generative model. Moreover, since both
ISIS-SCAD-$2$ and hierNet considered all the pairwise interactions
between $p$ predictor variables, they have computational complexity
$O(np^2)$ with $p=1000$ and need much more computational resources
compared with SIRI. On average, ISIS-SCAD-2 and hierNet are more than
$100$ times slower than SIRI (see Table~\ref{tabtime} for running time
comparison of different methods). While we can dramatically increase
the computational speed by using DC-SIS to screen variables before
applying more refined variable selection methods, relevant predictors
may be incorrectly filtered out by the DC-SIS procedure as shown by
DC-SIS-SCAD's higher false negative rates under Scenario $2.3$ of
Table~\ref{tabinteract1}.

As shown in Table~$9$ of \citet{Jiang2014}, both false positives and
false negatives increased when predictors were moderately or highly
correlated. DC-SIS-SCAD-$2$ performed the best under Scenario $2.1$,
since it assumes the same parametric form as the generative model, and
this assumption is important for selecting relevant predictors from
many correlated ones. When there were multiple pairwise interactions
(Scenario $2.3$), SIRI-AE outperformed DC-SIS-SCAD-$2$ as DC-SIS
falsely filtered out relevant predictors when their effects were weak.
When predictors were generated from the uniform distribution $\operatorname{Unif}(-2,
2)$ [Setting $4$ in Table~$9$ \citet{Jiang2014}], the performance of
SIRI was relatively robust under Scenarios $2.1$ and $2.2$ although the
normality assumption is violated. Under Scenario $2.3$, magnitudes of
interaction effects became much weaker when predictors were generated
from $\operatorname{Unif}(-2, 2)$ instead of the normal distribution. As a
consequence, both the FP- and FN-values increased for both SIRI-AE and
SIRI-CE compared with the normal case, and DC-SIS-SCAD-$2$ failed to
detect relevant predictors most of the time.

Under Scenario $2.4$ with three-way interactions, the computational
cost prevented us from directly applying ISIS-SCAD-$3$ to consider all
the three-way interaction terms. So we only compared the performance of
ISIS-SCAD-$3$ after variable screening using DC-SIS, that is,
DC-SIS-SCAD-$3$ in Table~\ref{tabinteract2}. DC-SIS-SCAD-$3$ performed
the best under different sample sizes as it assumed the form of the
underlying generative model. Among other methods, the performance of
SIRI-AE improved dramatically as sample size increased, whereas hierNet
had almost no improvement. When sample size $n=1000$, SIRI-AE was able
to select all relevant predictors with very low false positives.

Simulations in Scenarios 2.1--2.4 were generated under the same
model assumption as ISIS-SCAD-$k$ and DC-SIS-SCAD-$k$, which gives them
advantage in the comparison. Under Scenarios $2.5$ and $2.6$ of
Table~\ref{tabinteract3}, when the generative model goes beyond
multiplicative interactions, we can see that SIRI-AE and SIRI-CE
significantly outperformed other methods in detecting relevant
predictors with low false positives. In Table~\ref{tabtime}, we record
the average running time of different methods under Scenarios
2.1--2.3, $2.5$ and $2.6$. As expected, SIRI and DC-SIS-SCAD were
much more computationally efficient than hierNet and ISIS-SCAD, which
need to enumerate all the pairwise interaction terms.

\section{Real data examples}\label{secreal}

We applied SIRI to two real data examples. The first example studies
the problem of leukemia subtype classification with ultra-high
dimensional features. In the second example, we treat gene expression
level in embryonic stem cells as a continuous response variable, and
are interested in selecting regulatory factors that interact with DNA
and other factors to regulate expression patterns of genes.

\subsection{Leukemia classification}
For the first example, we applied SIRI-CE to select features for the
classification of a leukemia data set from high density Affymetrix
oligonucleotide arrays [\citet{Golub1999}] that have been previously
analyzed by \citet{Tibshirani2002} using a nearest shrunken centroid
method and by \citet{Fan2008} using a SIS-SCAD based linear
discrimination method (SIS-SCAD-LD). The data set consists of $7129$
genes and $72$ samples from two classes: ALL (acute lymphocytic
leukemia) with $47$ samples and AML (acute mylogenous leukemia) with
$25$ samples. The data set was divided into a training set of $38$
samples ($27$ in class ALL and $11$ in class AML) and a test set of
$34$ samples ($20$ in class ALL and $14$ in class AML).

\begin{table}
\caption{Leukemia classification results}
\label{tableuk}
\begin{tabular*}{\textwidth}{@{\extracolsep{\fill}}lccc@{}}
\hline
\textbf{Method} & \textbf{Training error} & \textbf{Test error} & \textbf{Number of genes}\\
\hline
SIRI-CE & $0/38$ & $1/34$ & \phantom{0}8 \\
SIS-SCAD-LD & $0/38$ & $1/34$ & 16 \\
Nearest shrunken centroid & $1/38$ & $2/34$ & 21 \\
\hline
\end{tabular*}
\end{table}

The classification results of SIRI-CE, SIS-SCAD-LD and nearest shrunken
centroids method are shown in Table~\ref{tableuk}. The results of
SIS-SCAD-LD and the nearest shrunken centroids method were extracted
from \citet{Fan2008} and \citet{Tibshirani2002}, respectively. SIRI-CE
and SIS-SCAD-LD both made no training error and one testing error,
whereas the nearest shrunken centroids method made one training error
and two testing errors. Compared with SIS-SCAD-LD, SIRI used a smaller
number of genes ($8$ genes) to achieve the same classification accuracy.

\subsection{Identifying regulating factors in embryonic stem cells}

The mouse embryonic stem cells (ESCs) data set has previously been
analyzed by \citet{Zhong2012} to identify important transcription
factors (TFs) for regulating gene expressions. The response variable,
expression levels of $12\mbox{,}408$ genes, was quantified using the RNA-seq
technology in mouse ESCs [\citet{Cloonan2008}]. To understand the ESC
development, it is important to identify key regulating TFs, whose
binding profiles on promoter regions are associated with corresponding
gene expression levels. To extract features that are associated with
potential gene regulating TFs, \citet{Chen2008} performed ChIP-seq
experiments on $12$ TFs that are known to play different roles in
ES-cell biology as components of the important signaling pathways,
self-renewal regulators, and key reprogramming factors. For each pair
of gene and one of these $12$ TFs, a score named transcription factor
association strength (TFAS) that was proposed by \citet{Ouyang2009} was
calculated. In addition, \citet{Zhong2012} supplemented the data set
with motif matching scores of $300$ putative mouse TFs compiled from
the TRANSFAC database. The TF motif matching scores were calculated
based on the occurrences of TF binding motifs on gene promoter regions
[\citet{Zhong2005}]. The data consists of a $12\mbox{,}408\times312$ matrix with
$(i,j)$th entry representing the score of the $j$th TF on the $i$th
gene's promoter region.

\begin{table}
\tablewidth=250pt
\caption{The ranks of $12$ known ES-cell TFs (among 312 predictors)
using SIRI-AE and COP}
\label{tabmotif}
\begin{tabular*}{250pt}{@{\extracolsep{\fill}}lcc@{}}
\hline
 & \multicolumn{2}{c@{}}{\textbf{Ranks}} \\[-6pt]
  & \multicolumn{2}{c@{}}{\hrulefill} \\
\textbf{TF names}& \textbf{SIRI-AE} & \multicolumn{1}{c@{}}{\textbf{COP}} \\
\hline
E2f1 & \phantom{0}1 & \phantom{0}1 \\
Zfx & \phantom{0}3 & \phantom{0}3 \\
Mycn & \phantom{0}4 & 10 \\
Klf4 & \phantom{0}5 & 19 \\
Myc & \phantom{0}6 & -- \\
Esrrb & \phantom{0}8 & -- \\
Oct4 & \phantom{0}9 & 11 \\
Tcfcp2l1 & 10 & 36 \\
Nanog & 14 & -- \\
Stat3 & 17 & 20 \\
Sox2 & 18 & -- \\
Smad1 & 32 & 13 \\
\hline
\end{tabular*}
\end{table}

\begin{figure}

\includegraphics{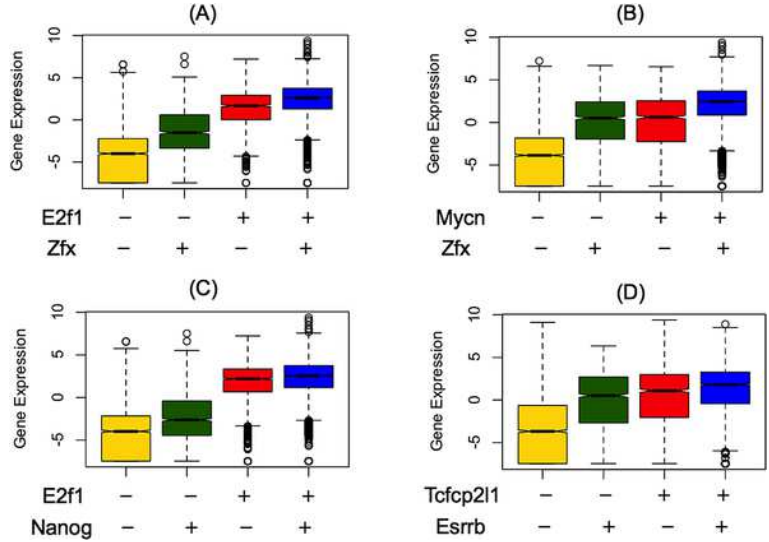}

\caption{The distribution of gene expression levels given the signs
(``$+$'' or ``$-$'') of TF motif matching scores. \textup{(A)} E2f1 and Zfx
(ranked among top $3$ by both SIRI and COP) show additive effects.
\textup{(B)}~Gene expression level is significantly lower when both Mycn (ranked 4
by SIRI and 10 by COP) and Zfx have negative scores. \textup{(C)} The matching
score of Nanog (ranked 14 by SIRI and missed by COP) has an effect on
gene expression only when E2f1 also has a negative score. \textup{(D)} Tcfcp2l1
(ranked 10 by SIRI and 36 by COP) and Esrrb (ranked 8 by SIRI and
missed by COP) have nonadditive effects in regulating gene expression.}
\label{figmotif}
\end{figure}

\citet{Zhong2012} reported that COP selected a total of $42$ predictors,
which include $8$ of the $12$ TFASs and $34$ of the $300$ TF motif
scores. Here, we used SIRI-AE to re-analyze the mouse ESCs data set and
selected $34$ predictors, which include all the $12$ TFASs and $22$ TF
motif matching scores. Relative ranks of the $12$ TFASs from SIRI-AE
and COP are shown in Table~\ref{tabmotif}. Among the top-$10$ TFs
ranked by SIRI-AE, $8$ of them are known ES-cell TFs. SIRI-AE is also
able to identify Nanog and Sox that are generally believed to be the
master ESC regulators but were missed in the results of COP. The ranked
list of 22 other TFs seleted by SIRI is given in \citet{Jiang2014}. A
further study of these TFs whose roles in ES cells have not been well
understood could help us better understand transcriptional regulatory
networks in embryonic stem cells.

In Figure~\ref{figmotif}, we illustrate combinatorial effects of
several identified TFs by plotting the distribution of gene expression
levels given the signs of a pair of TF motif matching scores. In
Figure~\ref{figmotif}(A), E2f1 and Zfx (ranked among top $3$ by both
SIRI and COP) have additive effects, that is, the combined effect of
two TFs is approximately equal to the sum of their individual effects
(which can be described by a linear model). The joint effects of TFs in
Figure~\ref{figmotif}(B), (C) and (D) show nonadditive patterns. For
example, in Figure~\ref{figmotif}(B), gene expression levels are
significantly lower when both Mycn and Zfx have negative matching
scores compared with other scenarios. A similar pattern is observed for
Tcfcp2l1 and Esrrb in Figure~\ref{figmotif}(D). Figure~\ref{figmotif}(C) shows that the effect of Nanog is only present when E2f1
has a negative matching score. As a result, COP, which is based on
linear combinations of TF matching scores, misses Nanog and Esrrb while
ranks Mycn and Tcfcp2l1 relatively lower. SIRI is able to identify
these TFs by capturing the nonadditive effects.

\section{Concluding remarks}\label{secconclusion}

We study the problem of variable selection in high dimensions from an
inverse modeling perspective. The contributions of the proposed
procedure that we named SIRI are twofold. First, it is effective and
computationally efficient in selecting relevant variables among a large
set of candidates useful for predicting the response, possibly through
complex interactions and other forms of nonlinear effects. Combined
with independence screening, SIRI can be used to detect complex
relationships in ultra-high dimensionality. Second, SIRI does not
impose any specific assumption on the relationship between the
predictors and the response, and is a powerful tool for variable
selections beyond linear models and for detecting variables with
unknown form of nonlinear effects. As a trade-off, SIRI imposes a few
assumptions on the distribution of the predictors. As demonstrated in
our simulation studies, SIRI has competitive performance when the
generative model is different from the inverse model assumption.
However, we found that SIRI is not very robust against extreme outliers
in values of the predictors. Data preprocessing, such as quantile
normalization, is advised when extreme outliers are spotted from
exploratory analysis. We have implemented the SIRI procedure using
programming language R, and the source code can be downloaded from \url
{http://www.people.fas.harvard.edu/\textasciitilde junliu/SIRI/} or requested from the
authors directly.

We have adopted an {ad hoc} rule to choose the slicing scheme in
SIRI. By allowing adaptive choices of slices based on observed data, we
are currently developing a dynamic programming algorithm to find the
optimal slicing scheme under a sliced inverse model. Theoretical
studies of such an algorithm, however, is more challenging and
delicate. Like other stepwise procedures such as linear stepwise
regression, SIRI may encounter issues that are typical to stepwise
variable selection methods as discussed in \citet{Miller1984}. When
relevant predictors have weak marginal effects but strong joint
effects, iterative sampling procedures such as Gibbs sampling could be
more powerful than stepwise procedures like SIRI. This motivates us to
further study the problem of variable selection from a full Bayesian
perspective.

The main goal of SIRI is to select relevant predictors with nonlinear
(including interaction and other second-order) effects on the response
without a specific parametric form. Without a specific parametric form,
however, it is impossible to precisely define what an ``interaction''
means. Interestingly, in many scientific problems, scientists often
cannot reach an agreement on what analytic form an interaction should
take even if they all agree that the interaction exists. As shown in
\citet{Zhang2007}, the inverse modeling approach as in na{\" i}ve Bayes
models (as well as in index models), we can finesse the interaction
definition problem by stating that the two predictors $X_1$ and $X_2$
have interactions if and only if their joint distribution conditional
on $Y$, {that is}, $ [X_1,X_2|Y ]$, cannot be factored
into the product of two marginal conditionals, {that is}, $
[X_1|Y ] [X_2|Y ]$. In order to be computationally
efficient, SIRI does not aim to pinpoint exactly which subsets ({e.g.}, pairs, triplets etc.) of variables are interacting sets, but
focuses on the overall set of predictors that may influence $Y$.
However, a follow-up study on the selected variables can provide
further information on which subsets of variables actually form an
``interaction clique'' in the sense of \citet{Zhang2007}.

Finally, inverse models are not substitutes of, but complements to,
forward models. When a specific form is derived from solid scientific
arguments, a forward perspective that treats the distribution of
predictors as a nuisance can be more powerful in building predictive
models. Depending on one's research questions and objectives, it may be
helpful to alternate between the two perspectives in analyzing and
interpreting data.

\begin{appendix}
\section*{Appendix: Proofs}\label{app}
\subsection{Proof of Theorem~\texorpdfstring{\protect\ref{thmsim}}{1} in
Section~\texorpdfstring{\protect\ref{seclrt}}{2.1}}\label{app2}
Given the set of relevant predictors indexed by $\mathcal{A}$ with size
$|\mathcal{A}|$ in model (\ref{eqsim}), we denote $B_\mathcal{A}=\operatorname
{Cov} (\mathbb{E}(\mathbf{ X}_{\mathcal{A}} |\break  S(Y)) )$, $W_\mathcal
{A}=\mathbb{E} (\operatorname{Cov}(\mathbf{ X}_{\mathcal{A}} | S(Y)) )$
and $\Omega_\mathcal{A}=B_\mathcal{A}+W_\mathcal{A}$. The corresponding
sample estimates are given by $\widehat{B}_{\mathcal{A}}$, $\widehat
{W}_{\mathcal{A}}$ and $\widehat{\Omega}_{\mathcal{A}} = \widehat
{B}_{\mathcal{A}}+\widehat{W}_{\mathcal{A}}$. To prove Theorem~\ref
{thmsim}, we will need the following lemma that is proved in \citet{Jiang2014}.

\begin{lemma}\label{lemma2}
Under the same conditions as in Theorem~\ref{thmsim}, for any set of
predictors indexed by $\mathcal{C}$, we let
$\widehat{\lambda}_i^{\mathcal{C}}$ be the $i$th largest eigenvalue of
$\widehat{\Omega}_{\mathcal{C}}^{-1}\widehat{B}_{\mathcal{C}}$ and let
$\lambda_i^{\mathcal{C}}$ be the $i$th largest eigenvalue of $\Omega
_{\mathcal{C}}^{-1}B_{\mathcal{C}}$. Then, for $0<\varepsilon<1$ and
$i=1,2,\ldots,q$, there exist positive constants $C_1$ and $C_2$ such that
%
\begin{eqnarray}
&&\operatorname{Pr} \Bigl(\max_{\mathcal{C} \subset\{1,2,\ldots,p\}} \bigl\llvert \log
\bigl(1-\widehat{\lambda}_i^{\mathcal{C}} \bigr)-\log \bigl(1-
\lambda _i^{\mathcal{C}} \bigr)\bigr\rrvert >\varepsilon \Bigr)
\nonumber
\\[-8pt]
\\[-8pt]
\nonumber
&&\qquad\leq 2p(p+1)C_1\exp{ \biggl(-C_2n
\frac{\tau_{\mathrm{min}}^4\varepsilon
^2}{64\tau_{\mathrm{max}}^2p^2} \biggr)},
\end{eqnarray}
where $\tau_{\mathrm{min}}$ and $\tau_{\mathrm{max}}$ are defined in
Condition~\ref{cond1}.
\end{lemma}

\begin{pf*}{Proof of Theorem~\ref{thmsim}}
Let $R_{\mathcal{C}} = \sum_{i=1}^q\log (1-\widehat{\lambda
}_i^{\mathcal{C}} )-\sum_{i=1}^q\log (1-\lambda_i^{\mathcal
{C}} )$. Then, according to Lemma~\ref{lemma2}, for $0<\varepsilon
<1$, there exist constant $C_1$ and $C_2$ such that
\[
\operatorname{Pr} \Bigl(\max_{\mathcal{C} \subset\{1,2,\ldots,p\}} \llvert R_{\mathcal{C}}
\rrvert > \varepsilon \Bigr) \leq2p(p+1)qC_1\exp{
\biggl(-C_2n\frac{\tau_{\mathrm{min}}^4\varepsilon^2}{64\tau_{\mathrm
{max}}^2p^2q^2} \biggr)}.
\]
Under Condition~\ref{cond2}, $p=o (n^{\rho} )$ with $2\rho
+2\kappa<1$, and for any positive constant~$C$,
\begin{eqnarray*}
&&\operatorname{Pr} \Bigl(\max_{\mathcal{C} \subset\{1,2,\ldots,p\}} \llvert
R_{\mathcal{C}}\rrvert > Cn^{-\kappa} \Bigr)
\\
&&\qquad\leq 2p(p+1)qC_1\exp{ \biggl(-C_2n^{1-2\kappa-2\rho}
\frac{\tau_{\mathrm
{min}}^4C^2}{64\tau_{\mathrm{max}}^2q^2} \biggr)} \rightarrow0
\end{eqnarray*}
as $n \rightarrow\infty$. For $j \notin\mathcal{C}$ and $d=|\mathcal{C}|$,
\begin{eqnarray*}
\widehat{D}_{j|\mathcal{C}} & =& -\sum_{i=1}^q
\log \bigl(1-\widehat {\lambda}_i^{d+1} \bigr) +\sum
_{i=1}^q\log \bigl(1-\widehat{
\lambda}_i^d \bigr)
\\
& = &-\sum_{i=1}^q\log \bigl(1-
\lambda_i^{d+1} \bigr) +\sum_{i=1}^q
\log \bigl(1-\lambda_i^d \bigr)-R_{[\mathcal{C}\cup\{j\}
]}+R_{\mathcal{C}}
\\
& = & \log \biggl(1+\frac{\operatorname{Var} (M_j )-\operatorname{Cov}
(M_j,\mathbf{X}_{\mathcal{C}} ) [\operatorname{Cov} (\mathbf
{X}_{\mathcal{C}} ) ]^{-1}\operatorname{Cov} (M_j,\mathbf
{X}_{\mathcal{C}} )^T}{\mathbb{E} (V_j )} \biggr)
\\
& & -R_{[\mathcal{C}\cup\{j\}]}+R_{\mathcal{C}},
\end{eqnarray*}
where
$M_j=\mathbb{E} (X_j|\mathbf{X}_{\mathcal{C}},S(Y) )$, $V_j=\operatorname
{Var} (X_j|\mathbf{X}_{\mathcal{C}},S(Y) )$, and $V_j$ is a
constant that does not depend on $\mathbf{X}_{\mathcal{C}}$ or $S(Y)$
under model (\ref{eqsim}).

When $\mathcal{C}^c \cap\mathcal{A} \neq\varnothing$, according to
definition of first-order detectable predictors, there exist $\kappa
\geq0$ and $\xi_0>0$ such that
\[
\max_{j \in\mathcal{C}^c \cap\mathcal{A}} \biggl[ \frac{\operatorname{Var}
(M_j )-\operatorname{Cov} (M_j,\mathbf{X}_{\mathcal{C}} ) [\operatorname
{Cov} (\mathbf{X}_{\mathcal{C}} ) ]^{-1}\operatorname{Cov}
(M_j,\mathbf{X}_{\mathcal{C}} )^T}{\mathbb{E} (V_j )} \biggr] \geq
\xi_0 n^{-\kappa}.
\]
Then, for sufficiently large $n$, there exists $j \in\mathcal{C}^c
\cap\mathcal{A}$ such that
\[
\log \biggl(1+\frac{\operatorname{Var} (M_j )-\operatorname{Cov} (M_j,\mathbf
{X}_{\mathcal{C}} ) [\operatorname{Cov} (\mathbf{X}_{\mathcal
{C}} ) ]^{-1}\operatorname{Cov} (M_j,\mathbf{X}_{\mathcal{C}}
)^T}{\mathbb{E} (V_j )} \biggr) \geq\frac{\xi_1}{2}n^{-\kappa}
\]
and
\[
\widehat{D}_{j|\mathcal{C}} \geq\frac{\xi_1}{2}n^{-\kappa} - \bigl(
\llvert R_{[\mathcal{C}\cup\{j\}]}\rrvert + \llvert R_{\mathcal{C}}\rrvert \bigr).
\]
Let $c=\frac{\xi_0}{4}$. Since
\[
\operatorname{Pr} \biggl(\max_{\mathcal{C} \subset\{1,2,\ldots,p\}} \llvert R_{\mathcal{C}}
\rrvert > \frac{c}{2} n^{-\kappa} \biggr) \rightarrow0,
\]
we have
\[
\operatorname{Pr} \Bigl( \min_{\mathcal{C}\dvtx \mathcal{C}^c\cap\mathcal{A} \neq
\varnothing} \max_{j \in\mathcal{C}^c\cap\mathcal{A}}
\widehat {D}_{j|\mathcal{C}} \geq c n^{-\kappa} \Bigr) \rightarrow1,
\]
as $n \rightarrow\infty$.

When variable $\mathcal{C}^c \cap\mathcal{A} = \varnothing$, for $j \in
\mathcal{C}^c \subset\mathcal{A}^c$, $M_j = \mathbb{E} (X_j|\mathbf
{X}_{\mathcal{C}},S(Y) )=\mathbb{E} (X_j|\mathbf{X}_{\mathcal
{C}} )$ is a linear combination of $\mathbf{X}_{\mathcal{C}}$ under
model (\ref{eqsim}), and
\[
\frac{\operatorname{Var} (M_j )-\operatorname{Cov} (M_j,\mathbf{X}_{\mathcal
{C}} ) [\operatorname{Cov} (\mathbf{X}_{\mathcal{C}} )
]^{-1}\operatorname{Cov} (M_j,\mathbf{X}_{\mathcal{C}} )^T}{\mathbb
{E} (V_j )} = 0.
\]
Thus,
\[
\widehat{D}_{j|\mathcal{C}} \leq \bigl(\llvert R_{[\mathcal{C}\cup\{j\}
]}\rrvert + \llvert
R_{\mathcal{C}}\rrvert \bigr)
\]
and
\[
\operatorname{Pr} \Bigl( \max_{\mathcal{C}\dvtx \mathcal{C}^c\cap\mathcal{A} =
\varnothing} \max_{j \in\mathcal{C}^c}
\widehat{D}_{j|\mathcal{C}} \geq Cn^{-\kappa} \Bigr) \leq\operatorname{Pr}
\biggl(\max_{\mathcal{C} \subset\{
1,2,\ldots,p\}} \llvert R_{\mathcal{C}}\rrvert \geq
\frac{C}{2}n^{-\kappa
} \biggr) \rightarrow0
\]
for any positive constant $C$ as $n \rightarrow\infty$.
\end{pf*}

\subsection{Proof of Theorem~\texorpdfstring{\protect\ref{thmsiri}}{2} in
Section~\texorpdfstring{\protect\ref{secaug}}{2.2}}\label{app5}

\begin{lemma}\label{lemma3}
Under the same condition as in Theorem~\ref{thmsiri}, for $0< \varepsilon
< 1$, there exist positive constants $C_1$ and $C_2$ such that
\[
\operatorname{Pr} \Bigl(\max_{\mathcal{C} \subset\{1,2,\ldots,p\}} \max_{j \in
\mathcal{C}^c}
\bigl\llvert \log\widehat{\sigma}_{j|\mathcal{C}}^2 -\log
\sigma_{j|\mathcal{C}}^2 \bigr\rrvert > \varepsilon \Bigr) \leq
\frac{p(p+1)}{2}C_1\exp \biggl(-C_2n\frac{\varepsilon^2}{p^2L^2}
\biggr)
\]
and
\begin{eqnarray*}
&&
\operatorname{Pr} \Biggl(\max_{\mathcal{C} \subset\{1,2,\ldots,p\}}
\max_{j \in\mathcal{C}^c} \Biggl\llvert \sum_{h=1}^Hs_h
\log \bigl[\widehat{\sigma}_{j|\mathcal{C}}^{(h)} \bigr]^2 -
\sum_{h=1}^Hs_h\log \bigl[
\sigma_{j|\mathcal{C}}^{(h)} \bigr]^2 \Biggr\rrvert >
\varepsilon \Biggr)
\\
&&\qquad\leq \frac{Hp(p+1)}{2}C_1\exp \biggl(-C_2n
\frac
{\varepsilon^2}{H^2p^2L^2} \biggr),
\end{eqnarray*}
where $L = \frac{4}{\tau_{\mathrm{min}}} (3 (\frac{\tau_{\mathrm
{max}}}{\tau_{\mathrm{min}}} )^{3/2}+1 )$, and $\tau_{\mathrm
{min}}$ and $\tau_{\mathrm{max}}$ are defined in Condition~\ref{cond1}.
\end{lemma}

\begin{pf*}{Proof of Theorem~\ref{thmsiri}}
We denote $R_{j|\mathcal{C}} = \log\widehat{\sigma}_{j|\mathcal{C}}^2
-\log\sigma_{j|\mathcal{C}}^2$ and
\[
\widetilde{R}_{j|\mathcal{C}}= \sum_{h=1}^Hs_h
\log \bigl[\widehat{\sigma }_{j|\mathcal{C}}^{(h)} \bigr]^2 -
\sum_{h=1}^Hs_h\log \bigl[
\sigma _{j|\mathcal{C}}^{(h)} \bigr]^2.
\]
According to Lemma~\ref{lemma3}, for $0<\varepsilon<1$, there exist $C_1$
and $C_2$ such that
\[
\operatorname{Pr} \Bigl(\max_{\mathcal{C} \subset\{1,2,\ldots,p\}} \max_{j \in
\mathcal{C}^c}
\llvert R_{j|\mathcal{C}} \rrvert > \varepsilon \Bigr) \leq\frac{p(p+1)}{2}C_1
\exp \biggl(-C_2n\frac{\varepsilon^2}{p^2L^2} \biggr)
\]
and
\[
\operatorname{Pr} \Bigl(\max_{\mathcal{C} \subset\{1,2,\ldots,p\}} \max_{j \in
\mathcal{C}^c}
\llvert \widetilde{R}_{j|\mathcal{C}}\rrvert > \varepsilon \Bigr) \leq
\frac{Hp(p+1)}{2}C_1\exp \biggl(-C_2n\frac{\varepsilon
^2}{H^2p^2L^2}
\biggr),
\]
where $L = \frac{4}{\tau_{\mathrm{min}}} (3 (\frac{\tau_{\mathrm
{max}}}{\tau_{\mathrm{min}}} )^{3/2}+1 )$.
Under Condition~\ref{cond2}, $p=o (n^{\rho} )$ and \mbox{$2\rho
+2\kappa<1$},
\begin{eqnarray*}
&&\operatorname{Pr} \Bigl(\max_{\mathcal{C} \subset\{1,2,\ldots,p\}} \max
_{j
\in\mathcal{C}^c} \llvert R_{j|\mathcal{C}}\rrvert > Cn^{-\kappa}
\Bigr)
\\
&&\qquad\leq \frac{p(p+1)}{2}C_1\exp{ \biggl(-C_2n^{1-2\kappa-2\rho}
\frac
{C^2}{L^2} \biggr)} \rightarrow0
\end{eqnarray*}
and
\begin{eqnarray*}
&&\operatorname{Pr} \Bigl(\max_{\mathcal{C} \subset\{1,2,\ldots,p\}} \max
_{j
\in\mathcal{C}^c} \llvert \widetilde{R}_{j|\mathcal{C}}\rrvert >
Cn^{-\kappa} \Bigr)
\\
&&\qquad\leq \frac{Hp(p+1)}{2}C_1\exp{ \biggl(-C_2n^{1-2\kappa-2\rho}
\frac
{C^2}{H^2L^2} \biggr)} \rightarrow0,
\end{eqnarray*}
for any positive constant $C$ as $n \rightarrow\infty$. We have
\begin{eqnarray*}
\widehat{D}^*_{j|\mathcal{C}} & = & \log\widehat{\sigma}_{j|\mathcal
{C}}^2
- \sum_{h=1}^Hs_h\log \bigl[
\widehat{\sigma}_{j|\mathcal
{C}}^{(h)} \bigr]^2
\\
&=& \log\sigma_{j|\mathcal{C}}^2 - \sum
_{h=1}^Hs_h\log \bigl[\sigma
_{j|\mathcal{C}}^{(h)} \bigr]^2+R_{j|\mathcal{C}}-\widetilde
{R}_{j|\mathcal{C}}
\\
& = & \log \biggl(1+\frac{\operatorname{Var} (M_j )-\operatorname{Cov}
(M_j,\mathbf{X}_{\mathcal{C}} ) [\operatorname{Cov} (\mathbf
{X}_{\mathcal{C}} ) ]^{-1}\operatorname{Cov} (M_j,\mathbf
{X}_{\mathcal{C}} )^T}{\mathbb{E} (V_j )} \biggr)
\\
& &{} +\log (\mathbb{E} V_j ) - \mathbb{E}\log
(V_j )+R_{j|\mathcal{C}}-\widetilde{R}_{j|\mathcal{C}},
\end{eqnarray*}
where
$M_j=\mathbb{E} (X_j|\mathbf{X}_{\mathcal{C}},S(Y) )$ and
$V_j=\operatorname{Var} (X_j|\mathbf{X}_{\mathcal{C}},S(Y) )$.

When $\mathcal{C}^c \cap\mathcal{A} \neq\varnothing$ and all the
relevant predictors indexed by $\mathcal{A}$ are stepwise detectable
with constant $\kappa$, then there exists $m \geq0$ such that $\bigcup_{i=0}^{m-1}\mathcal{T}_i \subset\mathcal{C}$ and $\mathcal{C}^c \cap
\mathcal{T}_m \neq\varnothing$. According to Definition~\ref{defstep},
there exist $j \in\mathcal{C}^c \cap\mathcal{T}_m$ and $\xi_1, \xi
_2>0$ such that either
\[
\frac{\operatorname{Var} (M_j )-\operatorname{Cov} (M_j,\mathbf{X}_{\mathcal
{C}} ) [\operatorname{Cov} (\mathbf{X}_{\mathcal{C}} )
]^{-1}\operatorname{Cov} (M_j,\mathbf{X}_{\mathcal{C}} )^T}{\mathbb
{E} (V_j )} \geq\xi_1 n^{-\kappa},
\]
that is, with sufficiently large $n$,
\[
\log \biggl(1+\frac{\operatorname{Var} (M_j )-\operatorname{Cov} (M_j,\mathbf
{X}_{\mathcal{C}} ) [\operatorname{Cov} (\mathbf{X}_{\mathcal
{C}} ) ]^{-1}\operatorname{Cov} (M_j,\mathbf{X}_{\mathcal{C}}
)^T}{\mathbb{E} (V_j )} \biggr) \geq\frac{\xi_1}{2}n^{-\kappa}
\]
or
\[
\log (\mathbb{E} V_j ) - \mathbb{E}\log (V_j ) \geq
\xi_2 n^{-\kappa}.
\]

Let $c = \min (\frac{\xi_1}{4},\frac{\xi_2}{2} )$. Therefore,
\begin{eqnarray*}
\widehat{D}^*_{j|\mathcal{C}} &\geq& \log \biggl(1+\frac{\operatorname{Var}
(M_j )-\operatorname{Cov} (M_j,\mathbf{X}_{\mathcal{C}} ) [\operatorname
{Cov} (\mathbf{X}_{\mathcal{C}} ) ]^{-1}\operatorname{Cov}
(M_j,\mathbf{X}_{\mathcal{C}} )^T}{\mathbb{E} (V_j )}
\biggr)
\\
& &{} + \log (\mathbb{E} V_j ) - \mathbb{E}\log
(V_j ) - \bigl(\llvert R_{j|\mathcal{C}}\rrvert +\llvert
\widetilde{R}_{j|\mathcal
{C}}\rrvert \bigr)
\\
&\geq& 2c n^{-\kappa} - \bigl(\llvert R_{j|\mathcal{C}}\rrvert +
\llvert \widetilde{R}_{j|\mathcal{C}}\rrvert \bigr).
\end{eqnarray*}
Since
\[
\operatorname{Pr} \biggl(\max_{\mathcal{C} \subset\{1,2,\ldots,p\}} \max_{j \in
\mathcal{C}^c}
\llvert R_{j|\mathcal{C}}\rrvert > \frac{c}{2} n^{-\kappa
} \biggr)
\rightarrow0
\]
and
\[
\operatorname{Pr} \biggl(\max_{\mathcal{C} \subset\{1,2,\ldots,p\}} \max_{j \in
\mathcal{C}^c}
\llvert \widetilde{R}_{j|\mathcal{C}}\rrvert > \frac{c}{2} n^{-\kappa}
\biggr) \rightarrow0,
\]
we have
\[
\operatorname{Pr} \Bigl( \min_{\mathcal{C}\dvtx \mathcal{C}^c\cap\mathcal{A} \neq
\varnothing} \max_{j \in\mathcal{C}^c\cap\mathcal{A}}
\widehat {D}^*_{j|\mathcal{C}}\geq c n^{-\kappa} \Bigr) \rightarrow1,
\]
as $n \rightarrow\infty$.

When $\mathcal{C}^c\cap\mathcal{A} = \varnothing$ under model (\ref
{eqaug}), for any $j \in\mathcal{C}^c$, $M_j = \mathbb{E}
(X_j|\mathbf{X}_{\mathcal{C}},\break S(Y) )=\mathbb{E} (X_j|\mathbf
{X}_{\mathcal{C}} )$, which is a linear combination of predictors
in $\mathbf{X}_{\mathcal{C}}$, and $V_j =  \operatorname{Var} (X_j|\mathbf
{X}_{\mathcal{C}},S(Y) ) = \operatorname{Var} (X_j|\mathbf{X}_{\mathcal
{C}} )$, which is a constant that does not depend on $\mathbf
{X}_\mathcal{C}$ or $S(Y)$. Then
\[
\frac{\operatorname{Var} (M_j )-\operatorname{Cov} (M_j,\mathbf{X}_{\mathcal
{C}} ) [\operatorname{Cov} (\mathbf{X}_{\mathcal{C}} )
]^{-1}\operatorname{Cov} (M_j,\mathbf{X}_{\mathcal{C}} )^T}{\mathbb
{E} (V_j )} = 0
\]
and
\[
 \log (\mathbb{E} V_j ) - \mathbb{E}\log
(V_j ) = 0.
\]
Thus,
\[
\widehat{D}^*_{j|\mathcal{C}} \leq\llvert R_{j|\mathcal{C}}\rrvert +\llvert
\widetilde{R}_{j|\mathcal{C}}\rrvert
\]
and
\[
\operatorname{Pr} \Bigl( \max_{\mathcal{C}\dvtx \mathcal{C}^c\cap\mathcal{A} =
\varnothing} \max_{j \in\mathcal{C}^c}
\widehat{D}^*_{j|\mathcal{C}} < Cn^{-\kappa} \Bigr) \rightarrow1,
\]
for any positive constant $C$ as $n \rightarrow\infty$.
\end{pf*}
\end{appendix}

\section*{Acknowledgements}
We thank Wenxuan Zhong for sharing the mouse embryonic stem cells data
set, Tingting Zhang for discussion on the COP procedure, Runze Li and
Wei Zhong for providing the R code for DC-SIS, Joseph K. Blitzstein and
Jessica Hwang for helpful suggestions on an earlier draft. The
authors are grateful to the Editor, the Associate Editor and three
referees for their insightful and constructive comments that helped to
greatly improve the presentation of the article.

\begin{supplement}[id=suppA]
\stitle{Supplement to ``Variable selection for general index models via
sliced inverse regression''}
\slink[doi]{10.1214/14-AOS1233SUPP} 
\sdatatype{.pdf}
\sfilename{aos1233\_supp.pdf}
\sdescription{We provide additional supporting materials that include
detailed proofs and additional simulation results.}
\end{supplement}



\printaddresses
\end{document}